\documentclass{aastex631}
\usepackage{enumerate}

\usepackage{lineno}

\shorttitle{Photoionization models of the M81 LINER}
\shortauthors{Li et al.}

\usepackage{graphicx}

\defcitealias{2022ApJ...928..111L}{Paper I}

\begin{document}

\title{CAHA/PPAK Integral-field Spectroscopic Observations of M81. II. Testing Photoionization Models in A Spatially-resolved LINER}
\author{Zongnan Li}
\email{lizn@bao.ac.cn}
\affiliation{National Astronomical Observatories, Chinese Academy of Sciences, A20 Datun Road, Chaoyang District, Beijing, 100101, China}
\author{Zhiyuan Li}
\affiliation{School of Astronomy and Space Science, Nanjing University, Nanjing 210023, China}
\affiliation{Key Laboratory of Modern Astronomy and Astrophysics, Nanjing University, Nanjing 210023, China}
\author{Rub\'{e}n Garc\'{\i}a-Benito}
\affiliation{Instituto de Astrof\'{\i}sica de Andaluc\'{\i}a (CSIC), P.O. Box 3004, 18080 Granada, Spain}
\author{Yifei Jin}
\affiliation{Research School of Astronomy \& Astrophysics, Australian National University, Canberra, 2611, Australia}

\begin{abstract}

The origin of the low-ionization nuclear emission-line region (LINER) prevalent in local galaxies and its relationship with supermassive black holes are debated for decades. 
We preform a comprehensive evaluation of traditional photoionization models against the circumnuclear ionized gas in M81, for which recent CAHA/PPAK integral-field spectroscopic observations reveal a LINER characteristic out to a galactocentric radius of $\sim$1 kpc. 
Constructed with the photoionization code {\sc cloudy}, the models have the novel aspect of their primary parameters being well constrained by extensive observations of a prototypical low-luminosity active galactic nucleus (LLAGN) and an old stellar bulge in M81.
Additionally, these models incorporate a reasonably broad range of uncertain nebular properties. 
It is found that the integrated photoionization  by the LLAGN and hot, low-mass stars distributed in the bulge can roughly reproduce the observed radial intensity distributions of the H$\alpha$, H$\beta$ and [N\,{\sc ii}] lines, with the bulge stars dominating the ionizing flux at radii $\gtrsim$200 pc. 
However, the models generally fail to reproduce a similarly declining profile of the [O\,{\sc iii}] line or an accordingly flat profile of the [O\,{\sc iii}]/H$\beta$ ratio.
This clearly points to a deficiency of ionizing photons in the outer regions despite an extended photoionization source. 
The discrepancy might be alleviated if much of the observed [O\,{\sc iii}] line arose from a bulge-filling, low-density gas surrounding a denser, H$\alpha$-emitting disk, or by a higher AGN luminosity in the recent past.
The case of M81 has important implications for the ionization mechanism of LINERs and low-ionization emission-line regions in general.

\end{abstract}
\keywords{Active galactic nuclei (16), Galaxy nuclei (609), Photoionization (2060), Galaxy spectroscopy (2171)}


\section{Introduction}

A supermassive black hole (SMBH), when present at the galactic nucleus, can have a diverse manifestation over a wide range of physical scales and wavebands, as the result of a continuous interplay between the SMBH and the host galaxy. The energy and momentum thus deposited into the circumnuclear environment and beyond, mediated primarily by radiation and mechanical outflows and collectively known as active galactic nucleus (AGN) feedback, are now understood as an indispensable element controlling the evolution of the host galaxy and the SMBH itself \citep{2012ARA&A..50..455F, 2013ARA&A..51..511K}. 
\\

The phenomenon of low-ionization nuclear emission-line regions \citep[LINERs;][]{1980A&A....87..152H} has been known for decades, 
and modern surveys establish that LINER is prevalent in nearby galaxies (see review by \citealp{2008ARA&A..46..475H}). 
As a class, LINERs are characterized by strong low-ionization forbidden lines of nitrogen (N), oxygen (O), and sulfur (S) with respect to the hydrogen (H) Balmer lines, arising from a nuclear volume ($\lesssim$ few hundred parsecs) coincident with the conventional narrow-line region (NLR) of Seyfert galaxies.
However, LINERs are not associated with a luminous AGN and exhibit only a moderate line luminosity compared to typical Seyfert nuclei. 
For this reason, and because of the early recognition that the prominent low-ionization lines cannot be generated by photoionization of ordinary OB stars, radiative shocks were first favored as the cause of LINERs \citep{1976ApJ...203L..49K, 1978MNRAS.183..549F, 1980A&A....87..152H}, due to their ability to both collisionally ionize/excite the post-shock gas and induce abundant ultraviolet (UV) photons that can be coupled to the pre-shock gas \citep[the so-called UV precursor;][]{1979ApJ...227..131S}. 
The physical origin of the shocks, however, is generally elusive, although it could be related to some mechanical activity of a central SMBH.  
The same SMBH, while accreting and radiating weakly compared to its counterpart in Seyfert nuclei, may still provide sufficient photoionization for at least a fraction of the LINERs \citep{1979ApJ...230..667K, 1983ApJ...264..105F, 1983ApJ...269L..37H}. Alternatively, a previously active SMBH may also serve as the central engine of LINERs \citep{1995ApJ...445L...1E}. 
This invoked the view that LINER is one manifestation of AGNs, and more importantly, added to the ever-growing evidence that SMBHs are prevalent in normal galaxies \citep{2013ARA&A..51..511K}.
\\

While photoionization by a low-luminosity AGN (LLAGN) remains a leading candidate for the 
ionization/excitation of LINERs, other sources, such as Wolf-Rayet stars \citep{1992ApJ...397L..79F, 2000PASP..112..753B}, hot low-mass evolved stars \citep{1994A&A...292...13B, 2000AJ....120.1265T, 2012ApJ...747...61Y, 2013A&A...558A..43S}, 
X-ray binaries \citep[XRBs;][]{2006A&A...460...45G, 2008ARA&A..46..475H}, 
hot X-ray emitting gas \citep{1989ApJ...345..153S, 1990ApJ...360L..15V}, and
cosmic rays \citep{1984ApJ...286...42F},  have also been considered.
Many, if not all, of these non-AGN mechanisms, share the property of being ubiquitous in normal galactic nuclei, thus each might have a non-negligible contribution to the observed nuclear emission lines. 
In particular, the hot low-mass evolved stars (HOLMES; \citealp{2011MNRAS.415.2182F}), which include post-asymptotic giant branch (pAGB) stars, central stars of planetary nebulae (PNe), and extreme horizontal branch (EHB) stars, are shown to be a viable mechanism of boosting low-ionization lines, due to a harder UV spectrum resulting from their significantly higher surface temperatures compared to OB stars.
In recent years, HOLMES has received renewed attention following the advent of integral field spectroscopic (IFS) observations of low-redshift galaxies,  
which revealed the existence of extended low-ionization emission-line regions \citep[LIERs;][]{2016MNRAS.461.3111B} on galaxy scales \citep{2010MNRAS.402.2187S, 2013A&A...558A..43S, 2016MNRAS.461.3111B,2019AJ....158....2B}.
LIERs are hard to explain with a centrally-concentrated ionizing source, i.e., an LLAGN \citep{2010MNRAS.402.2187S}. Spatially distributed ionizing sources, such as HOLMESs, appear more natural and appealing, although a critical test for this scenario necessarily involves more physical parameters than for AGN photoionization and remains to be done for individual galaxies.
\\

Understanding the ionization mechanism(s) of LINERs is certainly important for our general understanding of how and to what degree the energy of LLAGNs is coupled to the interstellar medium (ISM). 
However, it is generally challenging to distinguish the dominant ionization mechanism for a given LINER. 
This is because all aforementioned mechanisms can generate the characteristic low-ionization lines, and for a given mechanism, the conventional single-zone (i.e., a combination of constant physical parameters) models can in principle cover a sufficiently broad range of line intensity and intensity ratios. 
While the estimation of energy/photon budget can serve as a useful further test of a given mechanism \citep{1996ApJ...462..183H, 2009MNRAS.397..148L, 2010ApJ...711..796E}, this approach is indirect and often neglects details in the emission lines, such as their spatial and kinematic information. 
Relatively few studies have made use of the spatial information to constrain the ionization mechanism, due partly to the lack of spatially-resolved spectroscopy for a sizable sample of LINERs.
A notable recent example is the study of \cite{Molina_2018}, which used Hubble Space Telescope (HST) slit spectroscopy to trace the one-directional distribution of UV and optical emission lines in three LINERs out to a projected distance of $\sim$100 pc from the galactic center. 
These authors concluded that all three LINERs are best described by a composite ionization/excitation, in which an LLAGN photoionizes the gas within the central $\sim$20 pc, and shock excitation dominates at larger distances.
This demonstrates the potential of disentangling various ionization/excitation mechanisms given spatially-resolved information about LINERs.
On the other hand, \cite{Molina_2018} mainly relied on the conventional line ratio diagrams, 
widely known as the BPT diagram and variants \citep{Baldwin_1981, 1987ApJS...63..295V}, 
but did not attempt to quantify the spatial variation of the observed absolute line strengths with a detailed photoionization model.
\\

At a distance of 3.6 Mpc \citep{2001ApJ...553...47F}, where 1$\arcsec$ corresponds to 17.8 pc, the massive spiral galaxy M81 provides one of the best opportunities to study the ionization mechanism(s) of LINERs.
M81 hosts one of the nearest LLAGNs, known as M81*, 
which has a current bolometric luminosity of $\sim 2\times10^{41}$ erg s$^{-1}$ for a 7$\times 10^7$ M$_\odot$ SMBH \citep{2003AJ....125.1226D}. 
M81* has been extensively investigated by multi-wavelength observations, which reveal a variable, power-law X-ray continuum \citep{1996PASJ...48..237I}, a compact UV core with a featureless continuum \citep{1996ApJ...462..183H, 1997ApJ...481L..71D}, 
and a compact radio core with a one-sided jet \citep{2001MNRAS.321..767I}. 
The luminosity of M81* was found to vary with time at X-ray \citep{2001MNRAS.321..767I}, optical \citep{1996AJ....111.1901B}, centimeter \citep{1999AJ....118..843H} and millimeter wavelengths \citep{2007A&A...463..551S}, by factors of a few, over timescales of intraday to years. 
These observed properties of M81* are typical of LLAGNs and are well explained by a truncated thin disk in conjunction with an inner radiatively inefficient, advection-dominated accretion flow \citep[ADAF, e.g.][]{1999ApJ...525L..89Q, 2010ApJ...720.1033M, 2018MNRAS.476.5698Y}.
The broadband spectral energy distribution (SED) of M81* can also be well reproduced by a composite model consisting of a jet, an ADAF, and a truncated disk \citep{2014MNRAS.438.2804N}. 
More recently, a high-velocity, hot outflow from M81* was discovered by Chandra high-resolution X-ray spectroscopy \citep{2021NatAs.tmp..126S}, which is most likely driven by the hot accretion flow (i.e., the ADAF), providing further evidence of LLAGN activity. 
\\

Early optical spectroscopic observations of the M81 nucleus already led to the identification of both a type-1 Seyfert \citep{1981ApJ...245..845P}, due to the presence of broad Balmer lines, and a LINER \citep{1980A&A....87..152H}, which were later confirmed by \citet{1988ApJ...324..134F} with high-quality optical spectra. 
\citet{1996ApJ...462..183H}, combining HST high-resolution UV and optical spectra and ground-based optical spectra, further derived detailed properties of both the broad-line region (BLR) and the NLR within the central parsecs. 
It was found that the emission-line spectrum of the NLR broadly agrees with photoionization by an LLAGN, but the inferred SED of the LLAGN was only marginally sufficient to account for the required ionizing flux.  
A recent CAHA 3.5m telescope IFS observation presented by \citet[][hereafter paper I] 
{2022ApJ...928..111L} provides the first two-dimensional mapping of prominent optical emission lines from circumnuclear ionized gas in M81, out to a projected galactocentric radius of $\sim$ 1 kpc.
Based on the conventional line ratio diagnostics (mainly [N\,{\sc ii}]/H$\alpha$ and [O\,{\sc iii}]/H$\beta$), \citetalias{2022ApJ...928..111L} found that a LINER classification holds for the bulk of this circumnuclear ionized gas, but to what extent the LLAGN accounts for this LINER remains unclear. 
\\ 

A further advantage of M81 lies in a rich set of multiwavelength data that help quantify its stellar and interstellar components. 
In particular, the stellar bulge of M81 is found to be predominantly old ($\gtrsim$ 8 Gyr; \citealp{2000AJ....119.2745K}), indicating that the HOLMES, but not massive young stars, dominate the stellar photoionization.
Moreover, thanks to the proximity of M81, both the circumnuclear starlight and various phases of the ISM can be readily resolved at a scale of 1--10 pc, which is, for instance, $\sim$10 times better than achieved in the seminal Palomar survey of nearby galactic nuclei \citep{1997ApJ...487..568H}, and $\sim$100 times better than for the typical Sloan Digital Sky Survey galactic nuclei (at a redshift of $\sim$0.1; \citealp{2012ApJ...747...61Y}). 
Together with the well-constrained properties of the LLAGN, this makes M81 an excellent target to study the ionization mechanisms of the circumnuclear gas in a spatially-resolved manner.
\\

We take up the first step of such a task in this work, also the second of a series utilizing the CAHA 3.5m telescope IFS observations of the central $\sim$1 kpc of M81. 
Specifically, we confront the spatially-resolved circumnuclear emission lines with a set of models generated by the photoionization code {\sc cloudy} \citep{2017RMxAA..53..385F}, the input parameters of which are tightly constrained by empirical properties of the photoionizing sources and illuminated gas in M81. 
This allows us to conclude that photoionization, by the LLAGN, the stellar populations, or both, is highly unlikely to fully account for the ionization/excitation of the M81 LINER especially at distances $\gtrsim$100 pc from the galactic center.
In Section \ref{sec:2}, we outline the basic information of the IFS observation and the preparation of the emission line data, which were detailed in \citetalias{2022ApJ...928..111L}. 
The setup of the {\sc cloudy} photoionization models is described in Section \ref{sec:3}, highlighting assumptions on the LLAGN, stellar populations, and nebular properties. 
The model results in close comparison with the observed data are presented in Section \ref{sec:4}. 
Implications of this study are discussed in Section \ref{sec:5}, followed by a discussion in Section~\ref{sec:6}.
\\

\section{IFU observation and emission line data}

\label{sec:2}

The spectroscopic data of the central $\sim$1 kpc region of M81 is taken with the PPAK integral field unit (IFU) of the Potsdam Multi-Aperture Spectrograph (PMAS) instrument mounted on the 3.5 m telescope of the 
Hispanic Astronomical Center in Andalusia (CAHA) at Calar Alto. A three-pointing dithering scheme was adopted to ensure a full field-of-view (FoV) coverage of the spectral cube. We carried out V1200 observations with an exposure time of 3600 s per pointing (split into three individual exposures), while for the V500 setup we obtained observations with a total of 2100 s per pointing (split into seven individual exposures to avoid saturation).
A detailed description of the observations and data reduction process was presented in \citetalias{2022ApJ...928..111L}. Here we briefly summarize the main results that are most relevant to the 
present study. 
\\

\begin{figure}
\centering
\includegraphics[width= 7in]{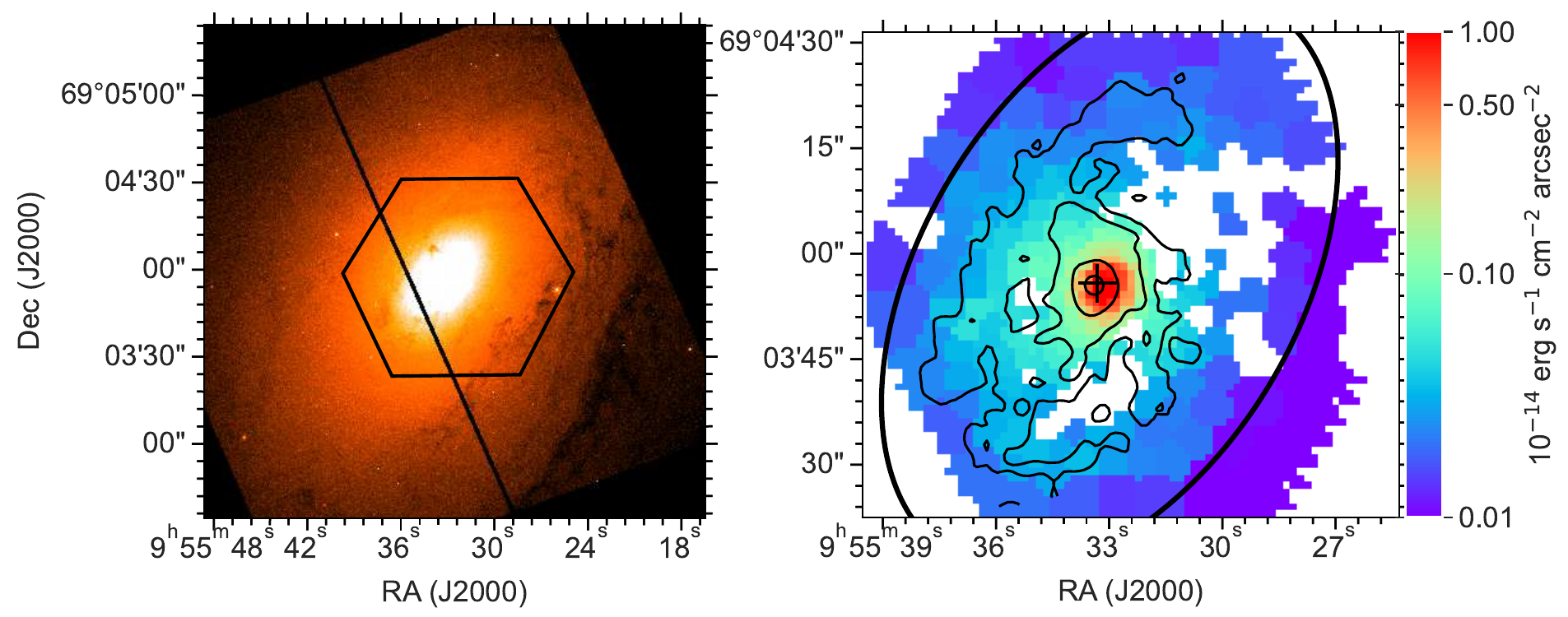}
\caption{$Lest$: HST WFC3/F438W image of the M81 bulge overlaid with the CAHA/PPAK FoV (black hexagon). $Right$: PPAK-measured H$\alpha$ line intensity after Voronoi tessellation binning \citepalias{2022ApJ...928..111L}. The H$\alpha$ intensity contours from an HST narrow-band image \citep{1997ApJ...481L..71D} delineate the nuclear spiral at a higher angular resolution. The black cross marks the position of M81*, and the ellipse outlines the region with a deprojected radius of 800 pc. \label{fig:hst_fov}}
\end{figure}

The PPAK IFU observations cover a hexagon FoV of 74$\arcsec~\times$ 64$\arcsec$ (Figure \ref{fig:hst_fov}), corresponding to a physical size of $\sim$ 1 kpc. The fiber size is $\sim 2.7\arcsec$, while the spaxel size is $1.0\arcsec$ ($\sim$ 18 pc) after dithering and regridding. 
The final wavelength coverage of the data ranges from 3700 $\rm\AA$ to 7300 $\rm\AA$, with a resolution of $\sim \rm 6~\AA$. The data reduction procedure follows the standard automatic pipeline of Calar Alto Legacy Integral Field spectroscopy Area survey \citep[CALIFA;][]{2012A&A...538A...8S, 2016A&A...594A..36S}. The reduced spectrum of each spaxel was fitted with the STARLIGHT software \citep{2005MNRAS.358..363C, 2011ascl.soft08006C}, following \cite{2017A&A...608A..27G}. The stellar spectrum was carefully modeled utilizing the synthesized spectrum of simple stellar populations (SSPs). The SSP-subtracted residual spectra contain prominent emission lines, including H$\alpha$($\lambda 6563$), H$\beta$($\lambda 4861$), [N\,{\sc ii}]$\lambda\lambda 6548, 6584$, [O\,{\sc ii}]$\lambda 3727$, [O\,{\sc iii}]$\lambda\lambda 4959,5007$ and [S\,{\sc ii}]$\lambda\lambda 6716, 6731$, which are standard diagnostics of the ionization and excitation of warm ($\sim 10^4$ K) gas. 
\\

\begin{figure}
\centering
\includegraphics[width= \columnwidth]{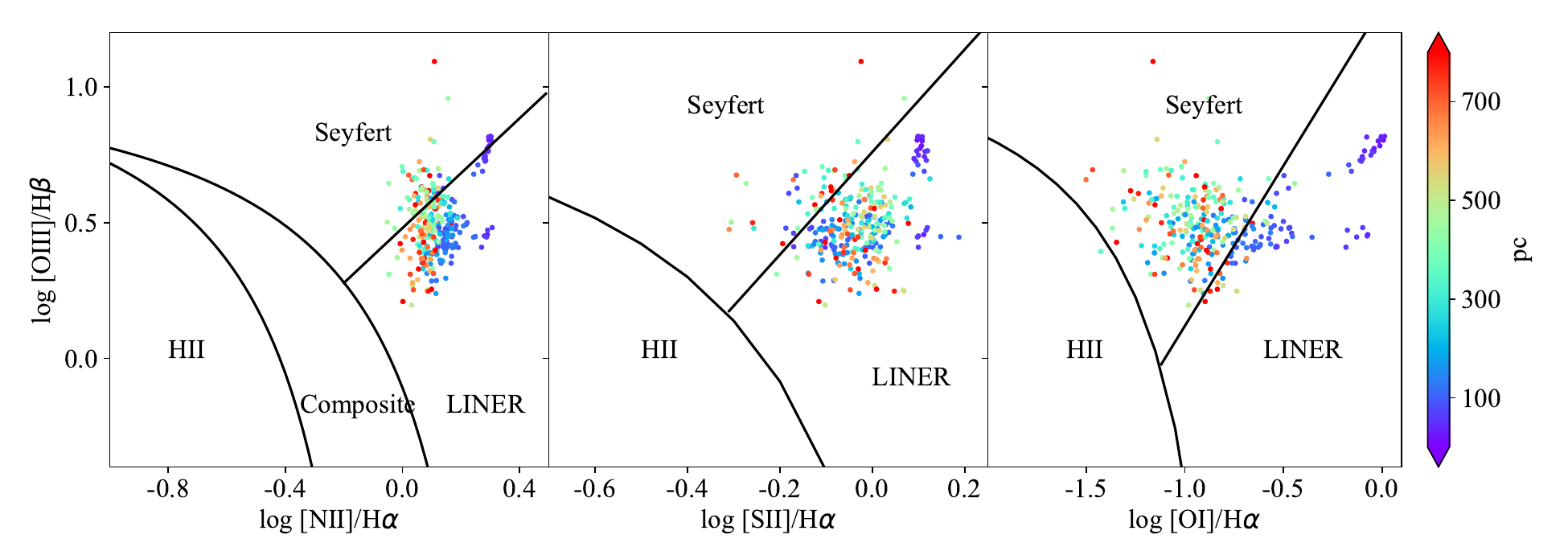}
\caption{The three traditional BPT diagrams plotting the flux ratios of emission line pairs [O\,{\sc iii}]$\lambda 5007$/H$\beta$ versus [N\,{\sc ii}]$\lambda 6584$/H$\alpha$, [S\,{\sc ii}]$\lambda\lambda 6716,31$/H$\alpha$, and [O\,{\sc i}]$\lambda 6300$/H$\alpha$. The data points are derived from each Voronoi bin and are color-coded by the deprojected
 distance from the galactic center. The dividing lines are from \cite{2006MNRAS.372..961K}, which divide the line ratios into Seyfert, LINER, HII, and composite regions. In all three diagrams, the circumnuclear region of M81 exhibits a mixture of LINER and Seyfert characteristics. 
\label{fig:BPT}}
\end{figure}

Due primarily to the relative faintness of the H$\beta$ line at the outer regions, we applied Voronoi tessellation binning \citep{2003MNRAS.342..345C} according to a target signal-to-noise ratio (S/N) of 7000 on the continuum for each bin, to maximize the useful information on the emission lines, at the cost of a degraded spatial resolution at the out regions. 
The resultant bin size increases with the galactocentric radius ($R$), from 18 pc within $R \sim 100$ pc to $\sim$120 pc at $R \sim$ 800 pc. 
These values are somewhat larger than the typical size of dense cold clouds probed by dust extinction against the bulge starlight, as seen in the HST broad-band images \citepalias{2022ApJ...928..111L},  but are representative of more diffuse ionized gas features. 
To obtain the intensity (and statistical error) of the emission lines, a single Gaussian was fitted to each line in the averaged spectra extracted from individual Voronoi bins.
H$\alpha$ and H$\beta$ were fitted with tied central velocity and velocity dispersion. 
A threshold of S/N $>$ 3 is adopted for all the lines of interest, which results in a valid fractional area of $\sim$85\%, as shown in Figure \ref{fig:hst_fov}, still ensuring an unbiased view of the circumnuclear region.
A map of the H$\alpha$ line intensity after binning is shown in Figure \ref{fig:hst_fov} as an illustration of the data.  
\\

For a better comparison with the models, we further corrected foreground extinction for the observed line intensities, based on a typical $A\rm_V$ of 0.3 toward the M81 center (\citealp{2000AJ....119.2745K}, \citetalias{2022ApJ...928..111L}) and the \cite{1989ApJ...345..245C} reddening curve, assuming $R\rm_V$ = $A\rm_V$/E(B -- V) = 3.1.  
The intensity ratios of a given line pair were also calculated for individual Voronoi bins, and the errors were obtained through error propagation. 
It is well-known that the nucleus of M81 exhibits a broad velocity component in the Balmer lines, which is a typical signature of the BLR of an AGN. In the CAHA/PPAK data, the BLR seems to extend up to a radius of $\sim 3''$ (53 pc), which is most likely due to PSF scattering effect \citepalias{2022ApJ...928..111L}. To minimize the contamination by the intrinsically compact BLR, we excluded the broad components (with a line width greater than $\sim$1000 km s$^{-1}$) of H$\alpha$ and H$\beta$.   
Therefore, all the line intensity and line ratios discussed below are supposed to trace the NLR and beyond. 

We then obtained the deprojected, azimuthally-average radial distribution of line intensities and line ratios measured from each Voronoi bin. 
The deprojected radius ($r$) of each bin was determined using the luminosity-weighted centroid and assuming that the gas clouds are primarily located in a plane aligned with the large-scale galactic disk. 
This is consistent with the bulk of the ionized gas found in arm-like features, collectively known as the {\it nuclear spiral}
(Figure~\ref{fig:hst_fov}).
We note that some fraction of the gas may be located outside the disk and spread in the bulge, the effect of which is discussed later. 
For the deprojection, we applied the disk parameters 
from \citetalias{2022ApJ...928..111L}, that is, position angle of the major-axis PA = $-30^\circ$ and inclination angle $i = 53^\circ$ (equivalent to an axis ratio of 0.6),  which are close to those derived by \cite{2011MNRAS.413..149S}. 
\\

Conventionally, the nature of the emission line regions can be distinguished with classical BPT diagrams using optical emission line ratio pairs, [O\,{\sc iii}]$\lambda 5007$/H$\beta$, [N\,{\sc ii}]$\lambda 6584$/H$\alpha$, [S\,{\sc ii}]$\lambda\lambda 6716,31$/H$\alpha$, and [O\,{\sc i}]$\lambda 6300$/H$\alpha$. As illustrated in Figure \ref{fig:BPT}, across the central kpc region of M81, the line ratios of individual bins show either a LINER or a Seyfert characteristic in all three diagrams, according to the classical definition of \cite{2006MNRAS.372..961K}. 
The small group of data points with the highest line ratios (in purple color) are from the innermost region after excluding the broad line components of the Balmer lines. 
Notably, those data points falling into the Seyfert region are mostly found at relatively large radii, 
which provide a critical test to the photoionization models (Section \ref{sec:4}).
\\

\section{Model description}
\label{sec:3}

We scrutinize the ionization mechanism(s) of the M81 LINER using the photoionization code {\sc cloudy}, version 17.03 \citep{2017RMxAA..53..385F}. {\sc cloudy} considers a slab of gas (hereafter referred to as a cloud) with specified geometry, density, and chemical composition, which is illuminated one-sided by a source or multiple sources with a specified spectral energy distribution (SED). The code then performs one-dimensional (1D) radiative transfer calculations under the conservation of mass, energy, and charge, to solve 
for the physical, chemical, and ionization states and emission spectrum of the gas. 
Main radiative heating and cooling processes are taken into account, which include free-free, free-bound, and bound-bound transitions of different elements. The physical, chemical, and ionization states in different layers of the cloud under local equilibrium are given as output.
\\

\begin{figure}
\centering
\includegraphics[width= 6in]{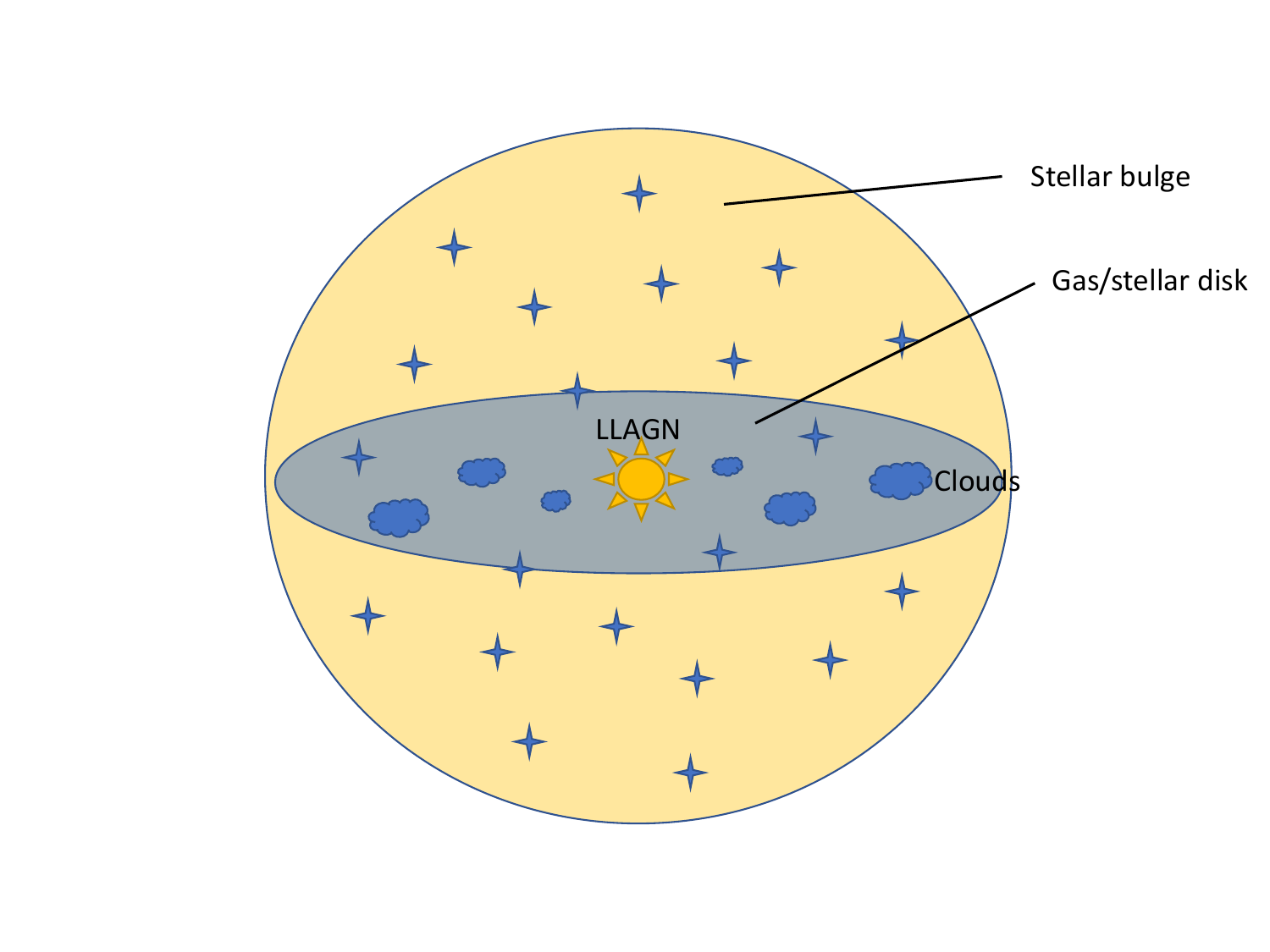}
\caption{A schematic view of the adopted geometry. The radially equally-spaced clouds are located in the same plane as defined by the stellar disk, illuminated by the central LLAGN and a combination of bulge+disk stars.
\label{fig:geo}}
\end{figure}

Our strategy is to place a series of equally-spaced clouds in the assumed gaseous disk plane of M81, which are illuminated one-sided by ionizing sources (a central AGN and/or stars distributed in the bulge/disk, see below) with specific SEDs and intensities. 
At any given radius, the model is similar to the conventional single-zone photoionization models, but collectively these modeled clouds, in response to the radiation field with a certain spatial variation, could provide the most stringent test to the assumed model.
An illustration of the geometry is shown in Figure \ref{fig:geo}. 
In a given {\sc cloudy} simulation, the total incident SED is assumed to be the sum of the individual SEDs, i.e., the central AGN or individual stars.
We also assume that the photoionization of a given cloud is not affected by the other clouds, effectively neglecting mutual obscuration by the clouds.
As for the line emission from the putative clouds, there are two general viewing geometry: When the cloud is on the far side of the disk (i.e., behind the AGN), the reflected emission is observed, whereas the transmitted emission dominates when the cloud is on the near side (i.e., in front of the AGN). In both cases, the emission lines are produced by the gas within the cloud, which are part of the diffuse radiation field in {\sc cloudy}. This radiation field is nearly isotropic and thus the resultant emission line properties from {\sc cloudy} models are independent of the viewing angle. 
\\

The above assumptions are akin to the 1D nature of the {\sc cloudy} code. In reality, the clouds may be illuminated in all directions. The number of ionization photons from a specific direction should be lower than in the assumed 1D case, which sums up ionizing photons from all directions. But this is more-or-less compensated by the fact that the emission lines from all illuminated sides of the cloud are actually counted, both in the model and in the observation. We have done a simple test by splitting the ionizing photons into two parts and rerunning the {\sc cloudy} simulation for a given cloud twice, each time injecting only one part of the ionization photons. It is found that the sum of the model-predicted emission line intensity from the two runs is quite close to that predicted by the integrated radiation field. Therefore  the modeled line intensities are not expected to be significantly biased by the assumed 1D geometry. 
A more realistic modelling taking into account the 3D geometry of the cloud, such as exemplified in \cite{2022ApJ...927...37J}, would be reserved for future investigation.  
\\

The resultant line intensities of e.g. H$\alpha$, [{\sc N\,ii}], and line ratios, e.g. [N\,{\sc ii}]/H$\alpha$, are then used to compare with the observed ones.
To match the resolution of the PPAK spectra, we specify the resolving power of the output spectra of {\sc cloudy} as 3000, which corresponds to a velocity resolution of $\sim$ 100 km s$^{-1}$.  
\\

\subsection{AGN model}
\label{sec:3.1}
The AGN is treated as a point source at the geometric center of the system. The SED of M81* is taken from \cite{2014MNRAS.438.2804N}, which fitted broadband data from a multi-wavelength campaign of M81* \citep{2008ApJ...681..905M} using either a jet-dominated (JD) model or an ADAF-dominated (AD) model. 
In reality, the X-ray emission of M81*, in particular the continuum, may be contributed by both the jet and the ADAF \citep{2021NatAs.tmp..126S}. For definiteness, here we adopt the jet-dominated SED in our simulations, except for one case in which the ADAF-dominated SED is adopted (Secction~\ref{subsec:AGNvar}). We note that the difference between the two models is only moderate, which is illustrated in the left panel of Figure \ref{fig:sed} comparing the two model SEDs. 
In particular, the ionizing UV ($\sim$100--912 \AA) photon flux, which is most responsible for the optical emission lines analyzed here, is only 1.15 times higher in the jet-dominated model than in the ADAF-dominated model. In this regard, our AGN SED is empirically well-constrained.
Notably, both the jet-dominated or ADAF-dominated models deviate substantially from the built-in AGN SED in {\sc cloudy} (also plotted in Figure \ref{fig:sed}), which exhibits a so-called {\it big blue bump} due to the classical accretion disk and is more suitable for AGNs accreting at a rate close to the Eddington limit.
\\

The total bolometric luminosity of M81* based on the adopted SED is estimated to be $L\rm_{bol} = 2.1 \times 10^{41}$ erg s$^{-1}$. 
We note that the intrinsic luminosity of M81* is known to have varied by a factor of a few over the past decades \citep{2007A&A...463..551S}, and we cannot rule out the possibility that it was much brighter 100--1000 yrs ago, a timescale relevant to the propagation of ionizing photons across the central kpc region. The consequence of a significantly enhanced AGN luminosity is tested and discussed in Section \ref{subsec:AGNvar}.
\\

\subsection{Stellar model}
\label{subsec:starmodel}
Besides the central LLAGN, stars can also play a significant role in ionizing and exciting the clouds. 
While young, massive stars can produce copious UV photons and are generally the leading candidate for stellar photoionization, the bulge of M81 is known to be deficient in recent star formation \citep{2000AJ....119.2745K}, which is consistent with the LINER/Seyfert classification (\citetalias{2022ApJ...928..111L}; Figure \ref{fig:BPT}). This is also supported by our own stellar population (SP) modeling based on the PPAK data (Z. Li et al. in preparation), which infers a typical stellar age of $\gtrsim 8$ Gyr in the central kpc region. Furthermore, a recent 3D radiative transfer modeling also suggests the dust heating fraction by young stars is $\lesssim 10\%$ within a radius of 3 kpc \citep{2020A&A...637A..24V}. Therefore, young stellar populations can be safely neglected.
On the other hand, the HOLMES \citep{2011MNRAS.415.2182F}, especially pAGB stars, EHB stars, and central stars of PNe, have long been considered potential sources of photoionization, due to a combination of high surface temperatures and large source numbers. 
We distinguish bulge and disk components for the old stars (Figure~\ref{fig:geo}), which can have different stellar ages and metallicity.
\\

We use the Flexible Stellar Population Synthesis (FSPS) model \citep{2009ApJ...699..486C, 2010ApJ...712..833C} integrated into Python by \cite{dan_foreman_mackey_2014_12157} to model the stellar populations in the bulge of M81. The stellar evolutionary tracks are from the MESA Isochrones \& Stellar Tracks \citep[MIST;][]{2016ApJ...823..102C, 2016ApJS..222....8D}, which are computed with the stellar evolution package Modules for Experiments in Stellar Astrophysics \citep[MESA v7503;][]{2011ApJS..192....3P, 2013ApJS..208....4P, 2015ApJS..220...15P}. 
The modeled SPs consist of single-age, single-metalicity stellar populations, spanning all metallicities and ages available. The initial mass function from \cite{2001MNRAS.322..231K} is adopted, with an upper and lower mass limit of 120 $M_\odot$ and 0.08 $M_\odot$, respectively. The age ranges from $10^{5-10.3}$ yr, with an interval of 0.05 dex, while the metallicity ranges from $-2.5$ to 0.5 on the logarithmic scale with respect to the solar metallicity $Z_\odot$ = 0.0142. 
The stellar evolution is computed all the way from the pre-main-sequence to the white dwarf phase. Thus, the HOLMES are naturally incorporated into our stellar model. 
\\

Specifically, we adopt a stellar population with an age of 8 Gyr and a metallicity of 1.5 $Z_\odot$ \citep{2000AJ....119.2745K} as the default bulge SP, which is roughly consistent with the PPAK stellar continuum data. For the disk SP, an age of 2 Gyr and a metallicity of $Z_\odot$ are assumed \citep{2000AJ....119.2745K}. 
The corresponding SEDs of these SPs for {\sc cloudy} are then generated using the publicly available code cloudyFSPS \citep{nell_byler_2018_1156412}. 
While the actual star formation history in M81 may potentially be more intricate than the assumed simple stellar population \citep{2020A&A...637A..24V}, the SED from all the HOLMES only weakly depends on the specific details of the star formation history, provided that ages exceed 10$^8$ yr.
\\

The amount and geometry of the bulge and disk SPs are determined as follows. We adopt the structural parameters of both the bulge and disk of M81 from the Spitzer Survey of Stellar Structure in Galaxies \citep[S$^4$G;][]{2010PASP..122.1397S, 2015ApJS..219....4S}, which provides decomposition parameters for 2352 nearby galaxies in $Spitzer$ IRAC 3.6/4.5 $\mu$m bands. 
For M81, the stellar mass of the bulge and the disk is $2.9 \times 10^{10} \rm ~M_\odot$ and $3.4 \times 10^{10} \rm ~M_\odot$, respectively \citep{2015ApJS..219....4S, 2015ApJS..219....5Q}. This is consistent with the total stellar mass of $1.9\times 10^{10} \rm~M_\odot$ integrated within our FoV (Z. Li et al. in preparation). 
To mimic the spatial distribution of ionizing stars in both the bulge and disk, we simulate 10$^6$ point sources for each component with 3D coordinates $P$($r$, $\theta$, $\phi$). 
The source number density at each spatial position is scaled with the decomposed IRAC 3.6 $\mu$m luminosity of the bulge and disk, respectively. The bulge stars are assumed to follow a deprojected S$\rm \acute{e}$rsic profile \citep{1968adga.book.....S} distribution with a S$\rm \acute{e}$rsic index of 3.56 and an effective radius of 1.38 kpc. The disk surface brightness profile is fitted with an exponential function with a scale length of 2.7 kpc and a scale height of 0.3 kpc in the S$^4$G catalog \citep{2015ApJS..219....4S}.
Here we assume that the bulge follows a spherical symmetry, and thus the other two coordinates, $\theta$, and $\phi$, are randomly distributed. The derived total integrated flux at a certain galactocentric radius is similar to that calculated using equation (9) given in \cite{2012ApJ...747...61Y}. 
\\

\begin{figure}
\centering
\includegraphics[width= 3.5in]{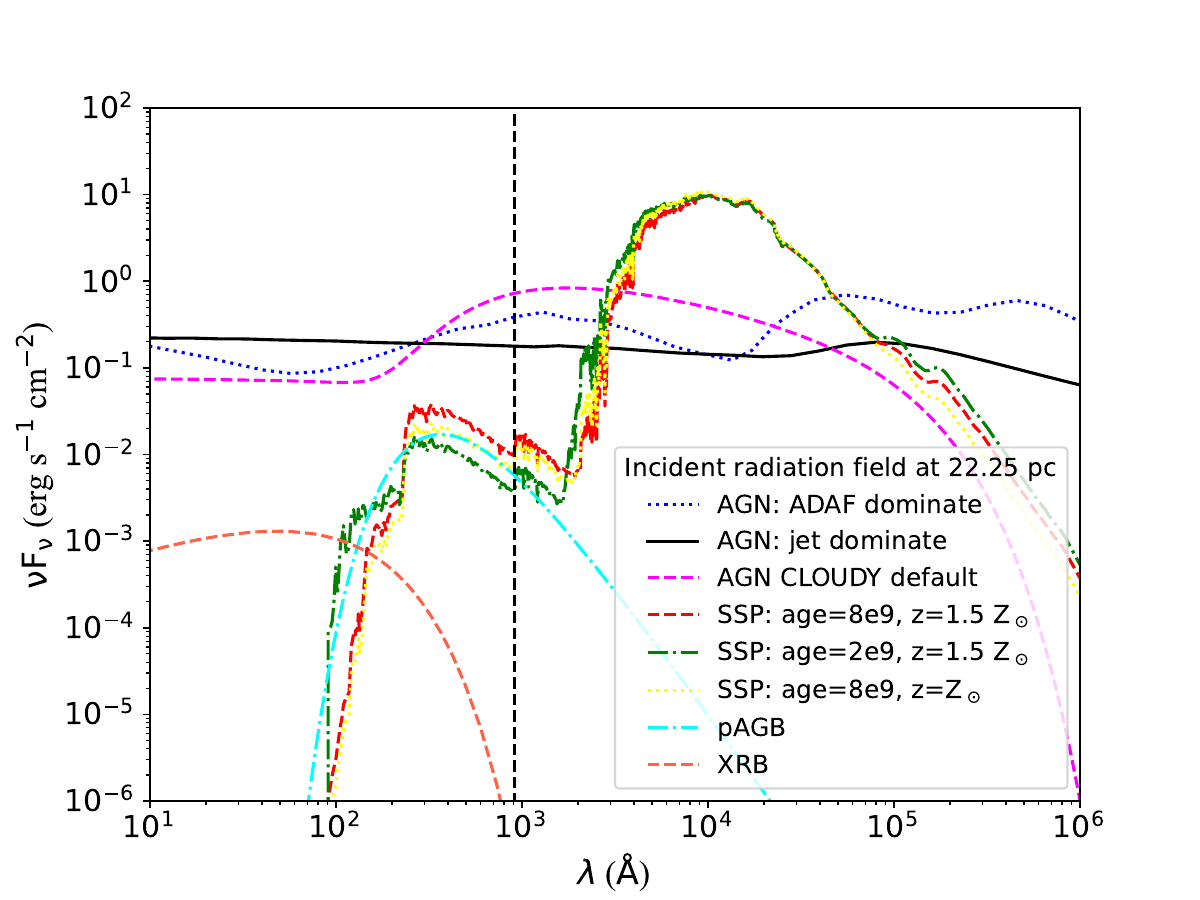}
\includegraphics[width= 3.5in]{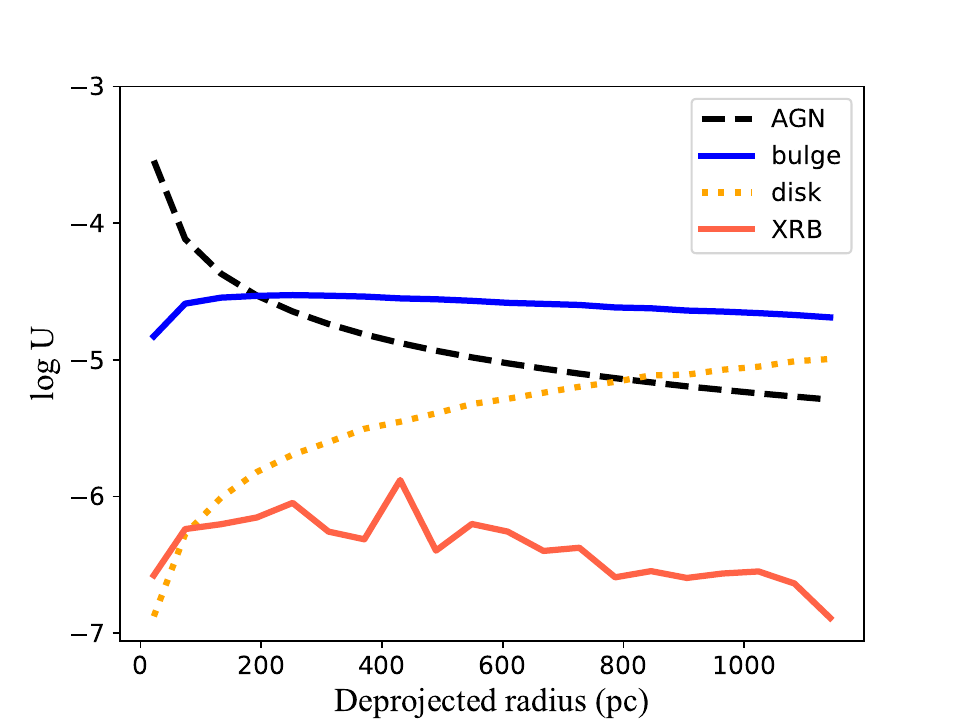}
\caption{$Left$: The adopted SED of the LLAGN (M81*) and stellar populations at $\sim$ 20 pc from the galactic center. 
The difference between the jet-dominated (solid black line) and ADAF-dominated (dotted blue line) models of M81* is small within the range of the ionizing UV band (100--912 $\rm\AA$). The default AGN model in {\sc cloudy} is also plotted as a dashed magenta line. 
The dashed red line is the default bulge SP, while the dash-dotted green and dotted yellow lines represent alternative bulge SPs with different ages and metallicities.
The dash-dotted cyan line shows the SED from 10000 pAGB stars each with a blackbody temperature of 10$^5$ K and a luminosity of 10$^3~L_\odot$, following the same deprojected S$\rm \acute{e}$rsic distribution as the bulge. 
The dashed orange line shows the SED of 200 XRBs following a $L^{-1}$ luminosity function and the deprojected S$\rm \acute{e}$rsic bulge distribution. The vertical black dashed line marks the Lyman limit at 912 $\rm\AA$. $Right$: The radial distribution of ionization parameter $U$,  by M81* (dashed red line), the default bulge SP (solid blue line), the disk SP (orange dotted line), and the XRBs (solid tomato line), respectively. 
\label{fig:sed}}
\end{figure}

The model-predicted SED of the bulge SP at a deprojected radius of $\sim$ 20 pc is shown in the left panel of Figure \ref{fig:sed} as an example. It can be seen that the UV flux at this radius is dominated by the LLAGN, while the bulge SP accounts for $\sim10\%$. Two additional bulge SPs, one with [age = 8 Gyr, Z = Z$_\odot$] and the other with [age = 2 Gyr, Z = 1.5 Z$_\odot$], are also plotted for comparison. 
It is evident that within the plausible ranges of age and metallicity, the three bulge SEDs are highly similar to each other, although the latter two SPs produce somewhat fewer ionizing photons. 
\\

The right panel of Figure \ref{fig:sed} further compares the radial distribution of the ionization parameter contributed by different sources, as predicted by {\sc cloudy} for the default setting of the illuminated clouds (Section~\ref{subsec:other}). Following \cite{1989agna.book.....O}, the dimensionless ionization parameter is defined as 
\begin{equation}
    U = \frac{\Phi(H)}{n(H)c},
\end{equation}
where $\Phi(H)$ is the surface flux of hydrogen-ionizing photons, $n(\rm H)$ is the total hydrogen density, and $c$ the speed of light.
Evident in this figure, the ionizing flux of the LLAGN has a $r^{-2}$ decline but dominates the overall photoionization within $\sim$200 pc. 
The bulge stars provide a much flatter profile of the ionizing flux, due to their extended distribution, 
which overtakes the LLAGN beyond $\sim$200 pc. 
We note that the small drop of the bulge ionization parameter in the innermost region is mainly due to a high gas density there.
The ionization parameter of the disk stars, on the other hand, increases gradually with radius, 
but is always lower than that of the bulge stars within the considered radial range. 
Therefore, we conclude that the old stellar populations in the bulge do have a substantial contribution to the photoionization in the central kpc region (especially beyond 200 pc) of M81. 
In the meantime, the disk has only a minor contribution. 
\\

In reality, most bulges are not truly spherical but are closer to triaxial with a mean short-to-long axis ratio of $\sim$0.5 \citep{2005ApJ...623..137V}. We have tested the 3D light distribution of the bulge with different triaxial shapes using the deprojection method presented in \cite{2021ApJ...914...45V}. It is found that with a short-to-long axis ratio of 0.5, the bulge starlight is more centrally concentrated, and the enclosed luminosity within 1 kpc is increased by $\sim$10\% on average, which, however, has little effect on the resultant radial distribution of the emission lines.
\\

\subsection{Other setups}
\label{subsec:other}
A total of 15 simulated clouds are assumed to lie in the same plane as defined by the galactic disk, with a galactocentric radius ranging from $\sim$ 20 pc to 850 pc at an interval of $\sim 60$ pc.
The inner boundary is chosen given the spatial resolution of the PPAK data, although the modeled SEDs described in the previous sections in principle remain valid at smaller radii. 
The azimuthal angle $\psi_{\rm cloud}$ is randomly assigned, as illustrated in Figure \ref{fig:geo}. 
The incident radiation field at the surface of a given cloud is calculated considering the relative position between the cloud and the simulated sources (the AGN or individual stars).
Here in {\sc cloudy}, we adopt the intensity case instead of the luminosity case to ensure a plane-parallel geometry and to facilitate comparison with the observed line intensity. We have also considered out-of-plane clouds with a fixed height of 300 pc from the disk. 
It is found that the incident intensity is lower by 0.2--0.5 dex compared to the case of in-plane clouds at the same projected radius $\lesssim 200$ pc, but is nearly unaffected at larger radii.
\\

We assume an open geometry since the covering factor of the clouds in the circumnuclear region is supposed to be small. Additionally, since this factor would have little influence on the line intensity, we do not specify it in the simulation. Regarding the volume filling factor, which quantifies the clumpiness of the clouds, previous studies suggest a value of 0.001 in the central 100 pc of M81 \citep{2011MNRAS.413..149S}. Nevertheless, modifying the filling factor only changes the thickness of the simulated clouds and has little influence on the overall output line intensity, 
which is supported by our tests using a filling factor of 0.001.
Thus the filling factor of each cloud is left as the default value of 1. 
The cosmic microwave background and cosmic ray background are not included in our models, since the former has a negligible effect in the UV band, whereas the influence of the latter heavily depends on the highly uncertain cosmic ray energy density, which will be reserved for future study.
\\

\begin{figure}
\centering
\includegraphics[width= 3.5in]{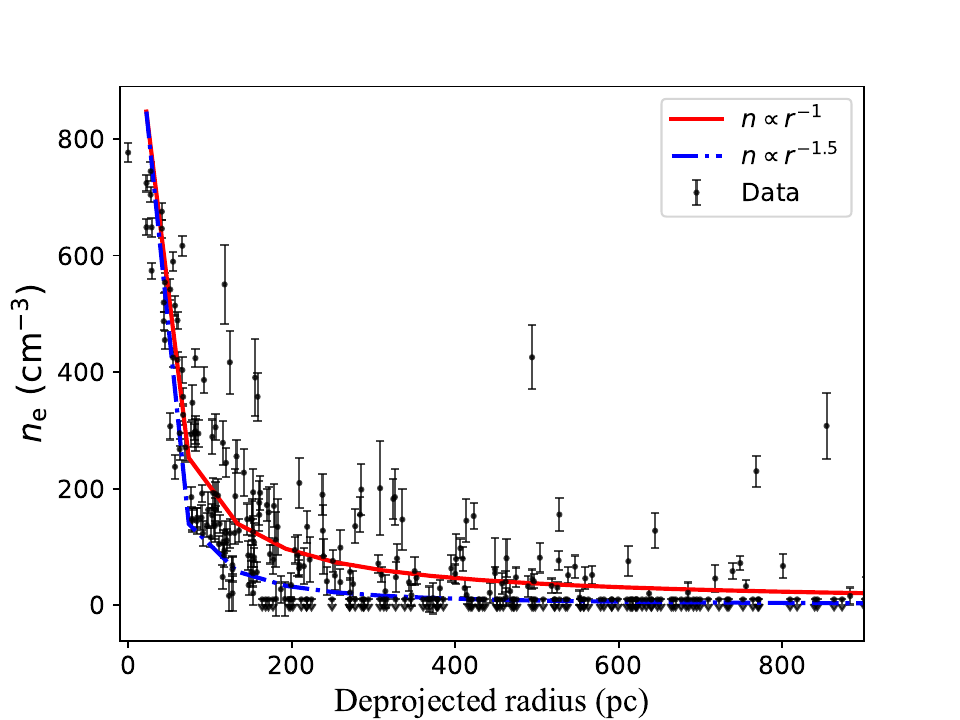}
\includegraphics[width= 3.5in]{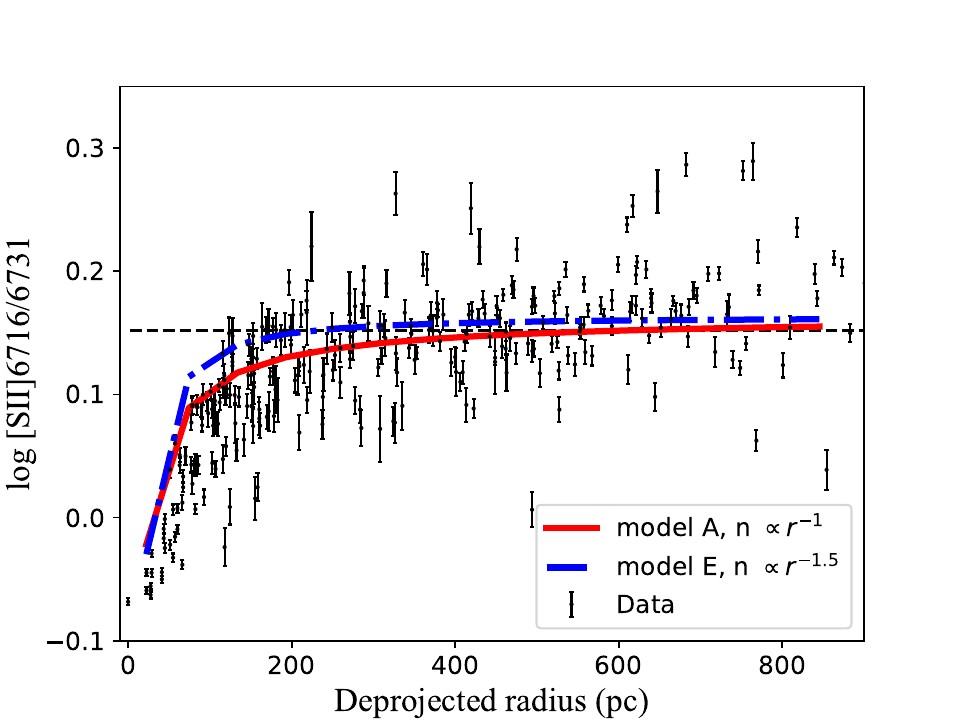}
\caption{$Left$: The radial distribution of electron density. The black points are computed with the observed [S\,{\sc ii}] doublet line ratio of each Voronoi bin. Some data points have their line ratio falling below the density-sensitive range, and are thus plotted as an upper limit at a value of 10 cm$^{-3}$. 
The default radial distribution of hydrogen density $n_{\rm H} \propto r^{-1}$ is shown as a solid red line. Another radial distribution, $n_{\rm H} \propto r^{-1.5}$, is shown by a dash-dotted blue line. $Right$: The observed [S\,{\sc ii}] doublet line ratio as a function of radius (black points), in comparison with model predictions based on the different hydrogen density profiles (Table \ref{tab:model}). The horizontal dashed line indicates the line ratio above which the resultant electron density is lower than 10 cm$^{-3}$, which is beyond the valid range of the [S\,{\sc ii}] doublet line ratio. 
\label{fig:ne}}
\end{figure}

An important parameter in the simulation is the hydrogen density $n_{\rm H}$. For a highly ionized gas, this is approximately equal to the electron density $n\rm_e$, which can be constrained by the observed intensity ratio of the [S\,{\sc ii}] doublet \citepalias{2022ApJ...928..111L}. 
The left panel of Figure \ref{fig:ne} shows the derived radial distribution of $n\rm_e$,
which has a relatively steep decline within the central 100 pc and becomes flattened at large radii. 
The values of $10^2-10^3\rm~cm^{-3}$ within the central 100 pc are
consistent with that inferred from the {\it Spitzer}/IRS [S\,{\sc iii}]$\lambda \rm 33 \mu m$/[S\,{\sc iii}]$\lambda \rm 18 \mu m$ intensity ratio \citep{2010ApJ...716..490S}.
We apply linear regression on log $n\rm_e$ using the least-squares fitting method of the {\it kmpfit} module\footnote{https://www.astro.rug.nl/software/kapteyn/kmpfit.html} in the Python package KAPTEYN, accounting for the measurement error.  
Since the [S\,{\sc ii}] doublet ratio is only sensitive to electron density within the range 10$^{1-4.5}\rm ~cm^{-3}$ \citep{1989agna.book.....O}, we exclude data points with log([S\,{\sc ii}]6716/[S\,{\sc ii}]6731) $> 0.15$ in the fitting. The resultant radial distribution is $n_e \propto r^{-1\pm 0.15}$, with a normalization of 847 cm$^{-3}$ at the radius of the innermost simulated cloud, 22.25 pc, which is also consistent with \citetalias{2022ApJ...928..111L}. Instead, if including the line ratios outside the sensitive range as upper or lower limits, and using the software Astronomy SURVival Analysis \citep[ASURV;][]{1985ApJ...293..192F} to fit the censored data, the resultant radial distribution would be closer to $n\rm_e \propto r^{-1.5\pm 0.1}$, which is the case of model F as discussed in section \ref{subsec:nebulavar}.
In our default model (Section~\ref{subsec:default}), we adopt for our simulated clouds a hydrogen density distribution following this fitted relation, rather than directly from the observed [S\,{\sc ii}] line ratio in view of the substantial statistical uncertainty.
As {\it a posteriori} check, the resultant electron density of the ionized surface layer of the simulated clouds is consistent with that derived from the [S\,{\sc ii}] line ratio, as is evident in the right panel of Figure \ref{fig:ne}. 
\\

The column density of the clouds can be constrained from the dust extinction map presented in \citetalias{2022ApJ...928..111L}, derived using an HST F438W image (Figure~\ref{fig:hst_fov}). 
The inferred hydrogen column density $N\rm_H$ is $\sim10^{20}~\rm cm^{-2}$ in most extinction features across the central kpc, but can reach $\sim$ 10$^{22}~\rm cm^{-2}$ in the darkest features. 
This is also in consistency with the molecular column density derived from IRAM 30m CO(1-0) observations, assuming a standard $N_{\rm H_2}/I_{\rm CO}$ conversion factor \citep{2007A&A...473..771C}, which is $\sim 5 \times 10^{20} \rm ~cm^{-2}$. 
We thus test three values, log $N\rm_H$ = 20, 21, and 22 in the simulation (Section \ref{subsec:nebulavar}), and adopt log $N\rm_H$ = 20 as the default condition. 
The resultant cloud thickness is $\sim$0.03--0.1 pc in the default case of log $N\rm_H$ = 20 and is larger for higher column densities. 
\\

We employ a constant pressure mode, assuming that the cloud illuminated by the radiation field is under total pressure equilibrium, which is shown to be more realistic and consistent with observations \citep[e.g.][]{2018ApJ...856...78A, 2023arXiv230512670Z}. 
In this case, the density is only defined in the illuminated cloud face, and evolves with cloud depth, as the temperature varies. Unless otherwise stated, we assume the cloud metallicity to be 1.5 times the Galactic ISM metallicity. We do not consider molecular gas and dust grains, because taking them into account has insignificant influences on the ionized gas properties discussed here (see further discussion in section \ref{subsec:nebulavar}). 
\\

\section{Model results}
\label{sec:4}

Several critical parameters, such as AGN luminosity, SP properties, and cloud properties, may affect the resultant line intensities. Therefore, we consider a total of twelve models (models A--L), each employing a different set of the main input parameters as summarized in Table \ref{tab:model}, to examine their effect on the line intensities and line ratios, as a function of deprojected radius from the galactic center. 
In the following, we first examine the physical properties and ionization states of the clouds in a default model and confront the model-predicted line properties with the observation (Section~\ref{subsec:default}). 
This is followed by the examination of additional models that employ either a different AGN luminosity (Section~\ref{subsec:AGNvar}), a different bulge SP (Section~\ref{subsec:bulgevar}), or different cloud properties (Section~\ref{subsec:nebulavar}).
\\

\subsection{The default model}
\label{subsec:default}
Our default model (model A) employs the sum of the LLAGN, the bulge SP (age = 8 Gyr, Z = 1.5 Z$_\odot$), and the disk SP (age = 2 Gyr, Z = Z$_\odot$) as the ionization source, a hydrogen density distribution following $r^{-1}$, 
a column density of $N\rm_H = 10^{20}~cm^{-2}$, and a 1.5 solar value for the gas metallicity.
\\

\begin{deluxetable*}{ccccccccc}
\tablecaption{{\sc cloudy} models. }
\tablenum{1}
\tablewidth{2pt}
\tablehead{
\colhead{Model} & 
\colhead{AGN SED\tablenotemark{a}} & 
\colhead{AGN luminosity} & 
\colhead{$t\rm_{SP (bulge)}$\tablenotemark{b}} & 
\colhead{$Z\rm_{SP (bulge)}$\tablenotemark{c}} & 
\colhead{Density profile\tablenotemark{d}} & 
\colhead{log $N\rm_H$} &
\colhead{$Z\rm_{cloud}$} &
\colhead{Note}\\
& &
\colhead{($\rm 2.1 \times 10^{41}~erg~s^{-1}$)} &
\colhead{(Gyr)} &
\colhead{($Z_\odot$)} &
&
\colhead{(cm$^{-2}$)} &
\colhead{($Z_\odot$)} &
}
\startdata
       A & JD & 1 & 10 & 1.5  & $\propto r^{-1}$ & 20 & 1.5 & AGN+SP \\
       B & AD & 1 & 10 & 1.5  & $\propto r^{-1}$ & 20 & 1.5 & AGN AD+SP\\
       C & JD & 10 & 10 & 1.5  & $\propto r^{-1}$ & 20 & 1.5 & AGN*10+SP\\
       D & JD & 1 & 2 & 1.5  & $\propto r^{-1}$ & 20 & 1.5 & SP age = 2 Gyr\\
       E & JD & 1 & 10 & 1  & $\propto r^{-1}$ & 20 & 1.5 & SP $Z=Z_\odot$\\
       F & JD & 1 & 10 & 1.5  & $\propto r^{-1.5}$ & 20 & 1.5 & $n \propto r^{-1.5}$\\
       G & JD & 1 & 10 & 1.5  & constant & 20 & 1.5 & $n = 1\rm ~cm^{-3}$\\
       H & JD & 1 & 10 & 1.5  & constant & 20 & 1.5 & $n = 10\rm ~cm^{-3}$\\
       I & JD & 1 & 10 & 1.5  & constant & 20 & 1.5 & $n = 100\rm ~cm^{-3}$ \\
       J & JD & 1 & 10 & 1.5  & $\propto r^{-1}$ & 21 & 1.5 & log $N\rm_H$ = 21\\
       K & JD & 1 & 10 & 1.5  & $\propto r^{-1}$ & 22 & 1.5 & log $N\rm_H$ = 22\\
       L & JD & 1 & 10 & 1.5  & $\propto r^{-1}$ & 20 & 1 & $Z_{\rm cloud} = Z_\odot$\\
\enddata
\tablecomments{$^a$Assumed AGN SEDs. JD stands for jet-dominated, and AD stands for ADAF-dominated. $^b$Age of bulge SP. $^c$Metallicity of bulge SP. $^d$Hydrogen density profile normalized to 847 cm$^{-3}$ at $r$ = 22.25 pc, except for the three constant density profiles. \label{tab:model}}
\end{deluxetable*}

\begin{figure}
\centering
\includegraphics[width=\textwidth]{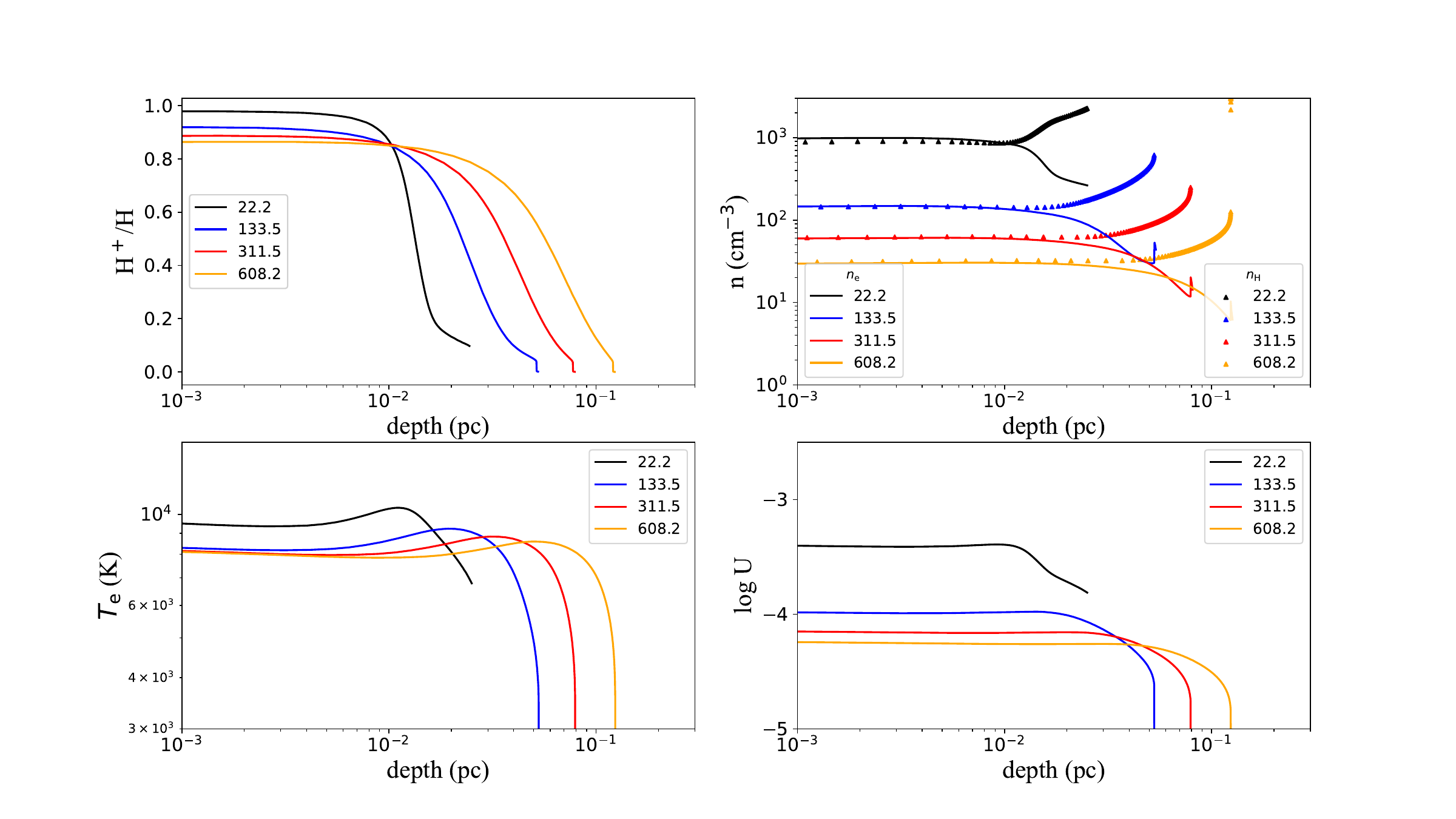}
\caption{Physical properties of individual clouds in the default model located at 22.2, 133.5, 311.5, and 608.2 pc, respectively, as a function of the depth from the illuminated surface. {\it Upper left}: the fraction of the H$^+$ ion; {\it Upper right}: electron and hydrogen density; {\it Lower left}: electron temperature; {\it Lower right}: ionization parameter. The outer edge of the cloud is determined by the stop column density, which is 10$^{20}~\rm cm^{-2}$ in the default model.
\label{fig:phys}}
\end{figure}

To examine the physical conditions of illuminated clouds at different deprojected radii, we plot in Figure \ref{fig:phys} the hydrogen ionization fraction, the electron density ($n_{\rm e}$), and hydrogen density ($n_{\rm H}$), electron temperature ($T_{\rm e}$), and ionization parameter ($U$) of selected clouds in the default model, as a function of depth from the illuminated surface. The model-predicted cloud thickness increases with the deprojected radius, ranging from 0.03 to 0.1 pc, due to an assumed constant hydrogen column density and a radially decreasing  hydrogen density. 
At all radii, the cloud surface is almost fully ionized in hydrogen, with the fraction of H$^+$ $\sim0.97$ at the innermost cloud and gradually decreasing to $\sim0.85$ at the outer clouds.
This is consistent with the values of $T\rm_e$ at the surface, ranging from $9.5\times10^3$ K at the innermost cloud to $8.0\times10^3$ K at the outer clouds. 
It is noteworthy that the electron temperature finds its maximum value not at the surface but at some depth into the cloud. 
This could be attributed to the soft ionizing photons being absorbed by metals closer to the ionizing source, which leaves hard ionizing photons penetrating into the cloud and producing more heat there. A similar behavior is found in high metallicity ($Z > 0.8 Z_\odot$) systems discussed in \cite{2019ARA&A..57..511K}.
The electron and hydrogen densities are close to each other in the ionized zone, and $n\rm_e$ gradually decreases with increasing cloud depth, while $n\rm_H$ increases in the meantime.
This is in accord with the decreasing H$^+$ fraction under the requirement of constant pressure.
The ionization parameter is log $U \sim -3.3$ in the innermost cloud at $\sim$20 pc, about one order of magnitude higher than in the cloud at $\sim$600 pc, beyond which $U$ drops only mildly, as already hinted in Figure~\ref{fig:sed}.
\\

\begin{figure}
\centering
\includegraphics[width=\textwidth]{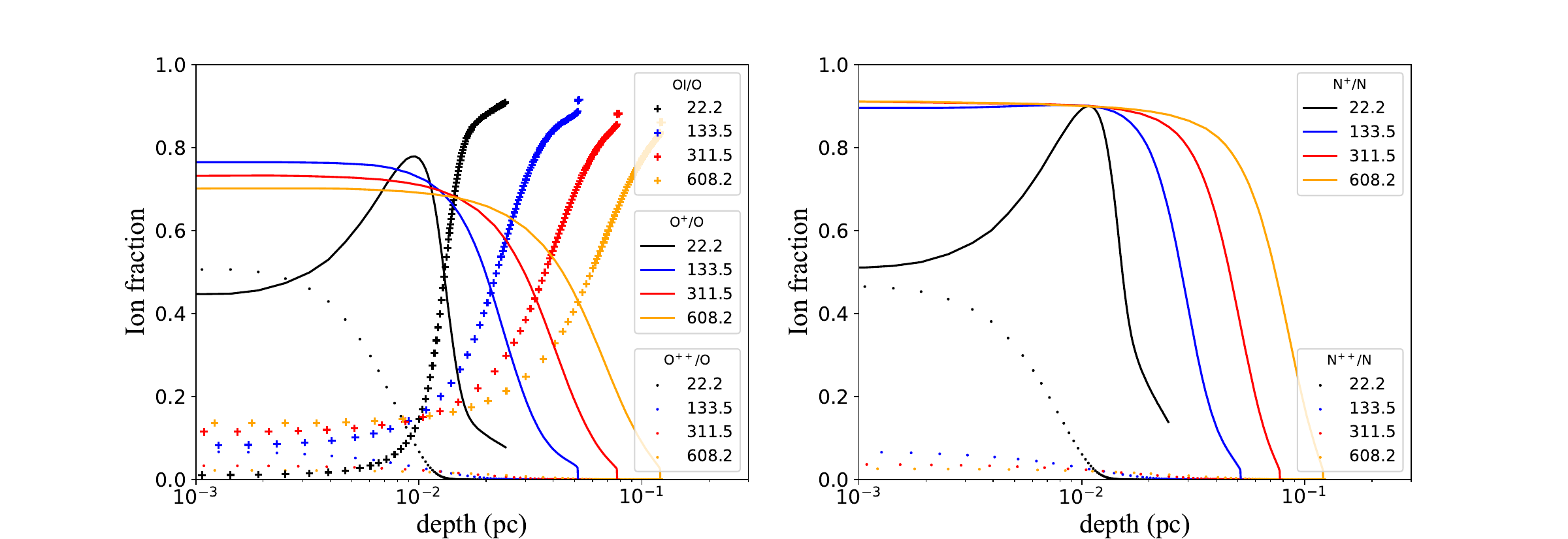}
\caption{The fraction of N and O ion species of individual clouds located at 22.2, 133.5, 311.5, and 608.2 pc in the default model, as a function of cloud depth. $Left$: Neutral oxygen (lines with crossed points), O$^{+}$ (solid lines) and O$^{++}$ (dotted lines) fraction. $Right$: N$^+$ (solid lines) and N$^{++}$ (dotted lines) fraction. The outer edge of the cloud is determined by the stop column density, which is 10$^{20}~\rm cm^{-2}$ in the default model.
\label{fig:ion_frac}}
\end{figure}

The behavior of the oxygen and nitrogen ion species are shown in Figure \ref{fig:ion_frac} left and right panel, respectively. 
The fraction of the doubly ionized species O$^{++}$ and N$^{++}$ is much higher in the innermost region than in the outer region ($\sim$50\% vs. $\lesssim$10\%). 
The trend is reversed for the fraction of the singly ionized species O$^+$ and N$^+$, which is lower in the innermost region than in the outer region ($\sim$50\% vs. $\sim$90\% near the surface). 
At the innermost cloud, both O$^+$ and N$^+$ exhibit a $\Lambda$-shaped behavior, peaking at a value of 80\% at a depth of $\sim$0.01 pc. 
Accordingly, a decreasing trend with cloud depth is seen in the O$^{++}$ and N$^{++}$ fraction. 
This can be understood as a depletion of sufficiently high-energy ionizing photons into the cloud. 
It is obvious that a high ionization parameter can result in more highly ionized O and N, while less singly ionized species. Therefore, with a sufficiently high ionization parameter, the singly ionized species can only exist beneath the illuminated surface of a cloud, also known as the {\it partially ionized zone}, where ionized and neutral species coexist. However, due to the moderate ionization field in our present case, most of the clouds are dominated by singly ionized species, as shown in Figure \ref{fig:ion_frac}. We have also examined the physical conditions and ionization structures of clouds in the other models. 
It is found that at a cloud increasing the ionization parameter increases the fraction of highly ionized species and substantially expands the ionized zone. 
On the other hand, clouds with a lower metallicity have an enhanced fraction of highly ionized species, a higher electron temperature, and a compressed partially ionized zone, which is mainly attributed to less cooling by metal lines at and near the cloud surface.
\\

\begin{figure}
\centering
\includegraphics[width= 7in]{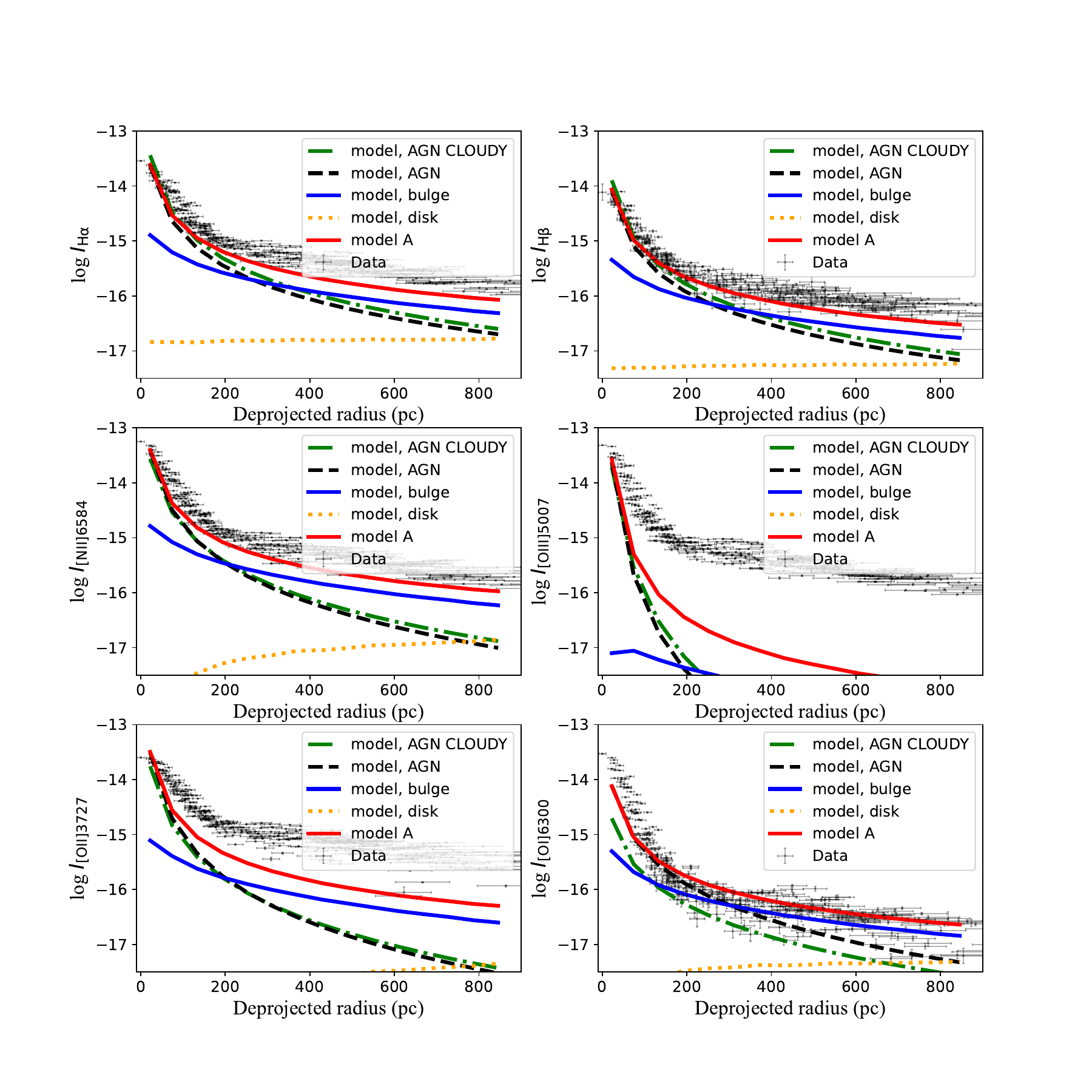}
\caption{The radial distribution of the observed line intensities (black points) and the simulated line intensities from the {\sc cloudy} models (colored lines) of most prominent emission lines: H$\alpha$, H$\beta$, [N\,{\sc ii}], [O\,{\sc iii}], [O\,{\sc ii}] and [O\,{\sc i}]. 
The units are $\rm erg~ s^{-1}~cm^{-2}~arcsec^{-2}$. The observed line intensities are derived from a single Gaussian fit to each Voronoi bin. The errorbar of the y-axis represents the standard deviation of the fitting results from 100 iterations of Monte Carlo simulations. The errorbar of the x-axis represents the range of the deprojected radius of each Voronoi bin. 
The dashed black line represents a {\sc cloudy} model with M81* as the only ionizing source, whereas the green solid line represents a similar model but using the built-in AGN SED of {\sc cloudy}.
The solid blue line represents a model with the 8-Gyr-old bulge SP as the only ionizing source, and the dotted orange line represents a model with the 2-Gyr-old disk SP as the only ionizing source. The combination of the AGN and the two SPs (i.e., model A) is shown by the solid red line. 
\label{fig:eml_agn_sp}}
\end{figure}

We now turn to comparison with the observation. Figure \ref{fig:eml_agn_sp} compares the radial distributions of observed (black points) and model-predicted (colored lines) line intensities, focusing on the H$\alpha$, H$\beta$, [N\,{\sc ii}], [O\,{\sc iii}], [O\,{\sc ii}] and [O\,{\sc i}] lines. 
All observed line intensity profiles exhibit a rapid decrease (by $\sim$ 1.5 dex) within a deprojected radius $\sim$ 200 pc, followed by a slow decline (by $\sim$ 0.5 dex) out to $\sim$800 pc. 
Remarkably, the default AGN+SP model (red solid curve) exhibits a similar trend with the observed profile for at least three lines: H$\alpha$, H$\beta$, and [N\,{\sc ii}].
This is an encouraging result, in the sense that almost all model parameters, especially those related to the ionizing sources, are empirically determined without any fine-tuning.
It is noteworthy that the observed H$\alpha$/H$\beta$ line ratio varies significantly across the FoV, with an average value of $\sim$3.6, which is significantly higher than the canonical Case B recombination value of 2.86, or 3.1 for AGN-dominated galaxies. This might be partially explained by intrinsic dust reddening, 
but we remind that such a moderately high Balmer decrement is not atypical of LINERs and could also be attributed to collisional excitation \citep{1980A&A....87..152H}.
\\

However, regarding the oxygen species, the model shows a substantial discrepancy with the observed profiles. This discrepancy is most significant in [O\,{\sc iii}], for which the model drops below the observed profile already at a radius of $\sim$100 pc, reaching a deficiency by up to two orders of magnitude at large radii. A similar behavior is found in [O\,{\sc ii}], although the discrepancy between the modeled and observed profiles is smaller. As for the [O\,{\sc i}] line, the model fits the observed profile well outside 150 pc, while predicting a lower intensity at smaller radii. 
The increasingly significant discrepancy from [O\,{\sc i}] to [O\,{\sc ii}] and to [O\,{\sc iii}] suggests an increasing deficiency of ionizing photons at progressively larger radii. 
\\

Figure \ref{fig:eml_agn_sp} also displays the line intensities resulting from the individual ionizing sources: the AGN (black dashed line), bulge (blue solid line), and disk (orange dotted line).   
It becomes clear that the AGN accounts for the central rise of the line profiles, but this source of ionizing photons alone declines too rapidly to account for the observed slow decline at radii beyond $\sim$200 pc.
The line intensities predicted by the {\sc cloudy} built-in AGN model are also plotted for comparison (dash-dotted green line). It is evident that this model in general predicts a higher intensity than the jet-dominated LLAGN SED, with the exception of the [O\,{\sc i}] line. This can be understood as the former has an on-average higher ionization parameter that tends to bring more O atoms into a higher ionization state. 
The bulge-only model driven by an extended stellar distribution, in principle, could be responsible for the observed slowly declining profiles at large radii.
However, the predicted line intensities fall short by $\gtrsim$ 0.5 dex even for the H$\alpha$, H$\beta$, and [N\,{\sc ii}] lines. If the inclination angle of the gaseous disk were in fact smaller than adopted here, the discrepancy can be somewhat alleviated (see further discussion in Section~\ref{sec:5}).
Finally, the disk-only model, as expected, has a negligible contribution for all lines at all radii of interest. 
\\

\begin{figure}
\centering
\includegraphics[width= 3.5in]{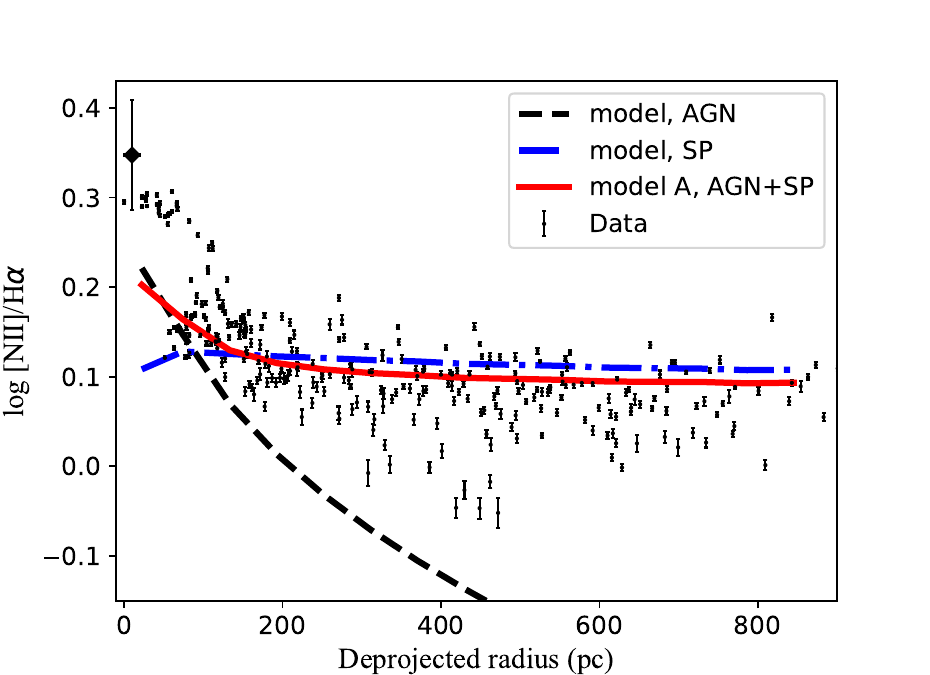}
\includegraphics[width= 3.5in]{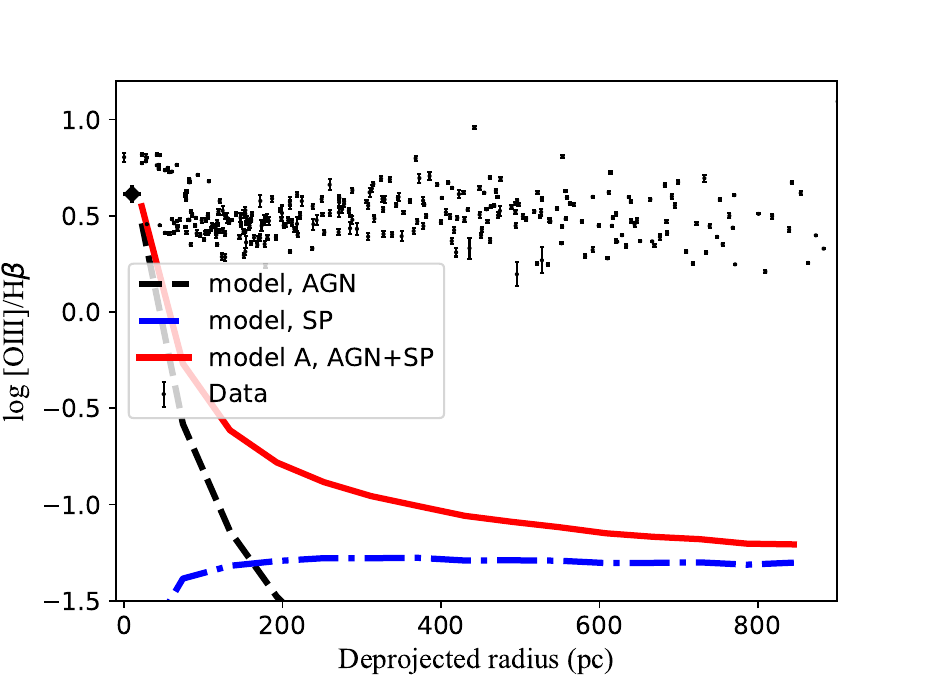}
\caption{
The radial distribution of observed [N\,{\sc ii}]/H$\alpha$ ({\it Left}) and [O\,{\sc iii}]/H$\beta$ ({\it Right}) line ratios are shown as black points.
For clarity, errorbars are not displayed for the x-axis. 
The line ratio measurements of ground-based spectroscopic observations extracted from an aperture of $\sim$4$''\times2''$ \citep{1996ApJ...462..183H} are also plotted as a black diamond. 
The AGN, SP (bulge plus disk), and AGN+SP models are shown by the black dashed, blue dash-dotted, and red solid lines, respectively.
\label{fig:ratio_agn_sp}
}
\end{figure}

Figure \ref{fig:ratio_agn_sp} plots the line ratios of [N\,{\sc ii}]/H$\alpha$ and [O\,{\sc iii}]/H$\beta$, which together provide the conventional BPT-diagram diagnostics. The log\,[N\,{\sc ii}]/H$\alpha$ ratio declines from a value of $\sim$0.3 in the innermost region to around 0.1 at a radius of $\sim$150 pc, then remains flat (albeit with substantial scatter) outwards. This behavior is not reproduced by either the AGN (black dashed line) or the SP (blue dash-dotted line) alone, but interestingly, the model of AGN+SP (red solid line) shows a similar behavior, although the rise predicted by the model in the innermost region has a smaller amplitude. 
The observed log\,[O\,{\sc iii}]/H$\beta$ ratio has a similar behavior, which declines from a value of $\sim$0.8 in the innermost region to $\sim 0.5$ at $\sim$150 pc, and remains flat outwards (with a scatter comparable to that in [N\,{\sc ii}]/H$\alpha$). 
However, the default AGN+SP model fails to explain this flat profile, nor can the AGN-only or SP-only models. 
For comparison, the line ratio measurements derived from ground-based spectroscopic observations of a $\sim$4$''\times2''$ aperture taken from Palomar and Lick observatories at multiple epochs \citep{1996ApJ...462..183H} are also plotted in the figure as black diamonds. These values are comparable, but not identical, to our measurements at the innermost few pixels. The difference might be explained by a combined effect of different apertures and different observing epochs.
\\

\subsection{Varying AGN property}
\label{subsec:AGNvar}
\begin{figure}
\centering
\includegraphics[width=3.5 in]{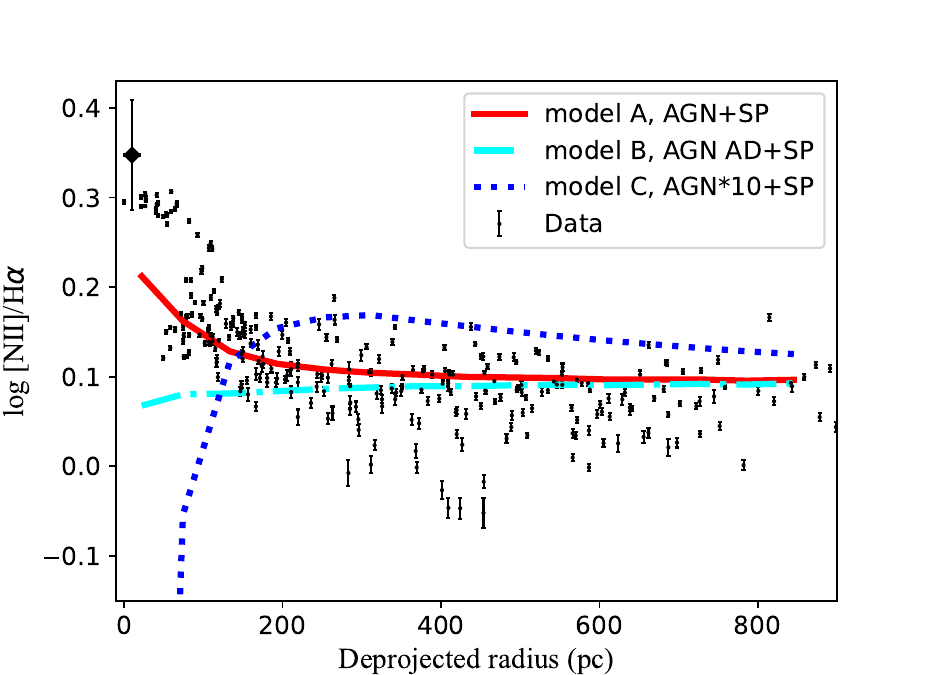}
\includegraphics[width=3.5 in]{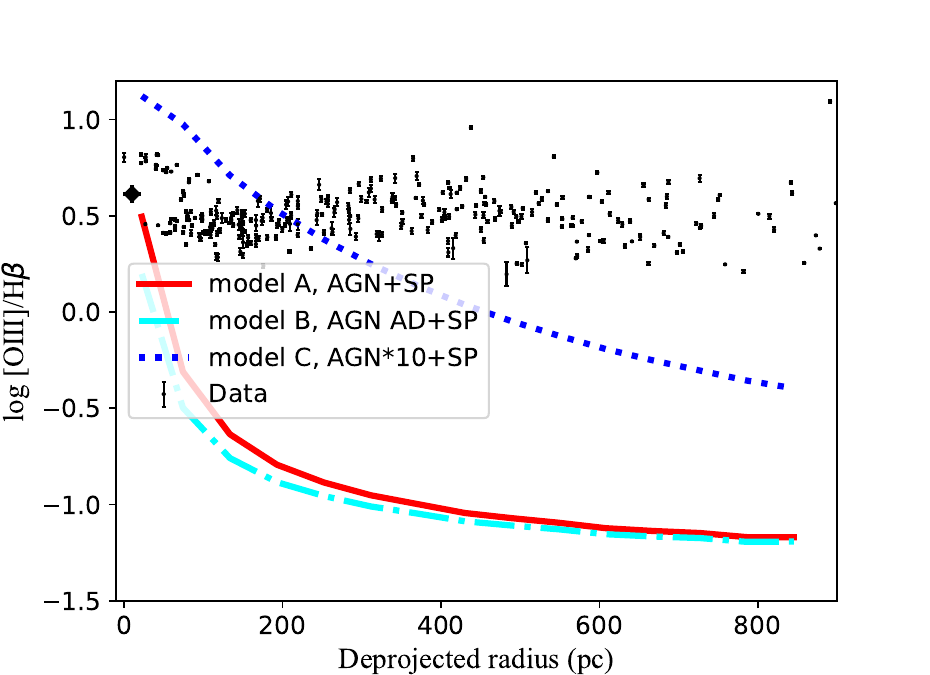}
\caption{The radial distribution of the observed [N\,{\sc ii}]/H$\alpha$ ($Left$) and [O\,{\sc iii}]/H$\beta$ ($Right$) line ratios, in comparison with models with different AGN luminosities. The default model (A) is shown by a solid red line, while model B with the ADAF-dominated SED is shown by the dash-dotted cyan line, and model C with a 10 times higher AGN luminosity is shown by the dotted blue line.
\label{fig:high_AGN}}
\end{figure}

To assess the difference between the two LLAGN SEDs (Section~\ref{sec:3.1}), we test a model (model B) that incorporates an ADAF-dominated SED, while keeping all other configurations identical to model A. The resulting [N\,{\sc ii}]/H$\alpha$ and [O\,{\sc iii}]/H$\beta$ ratios are depicted in Figure \ref{fig:high_AGN}, alongside model A and the data. Overall, model B predicts lower line ratios in the central region compared to model A, primarily due to the lower ionizing photon flux provided by this model. 
\\

Since M81*, and LLAGNs in general, are known to exhibit substantial intrinsic variability, we test a model (model C) in which the AGN bolometric luminosity is 10 times higher than in model A, but with the same SED shape. 
In this case, the bolometric luminosity is still $\lesssim10^{-3}$ of the Eddington luminosity of M81*, validating the use of a jet-ADAF-type SED \citep{2014MNRAS.438.2804N}.
The same SP component is applied as in model A, as well as the other model parameters.
The resultant [N\,{\sc ii}]/H$\alpha$ and [O\,{\sc iii}]/H$\beta$ ratios of model C are shown in Figure \ref{fig:high_AGN}, in comparison with model A and the data. The discrepancy between the observed and predicted [O\,{\sc iii}]/H$\beta$ ratio beyond 200 pc is much reduced, but still significant, in model C. 
Moreover, this comes at the price of a significantly higher [O\,{\sc iii}]/H$\beta$ ratio within 200 pc. 
Besides, the resultant [N\,{\sc ii}]/H$\alpha$ ratio of model C exhibits a steep decline within 200 pc, which is opposite to the observed trend. This is mainly because a significant fraction of the N atoms have been ionized to a higher state, N$^{++}$, under the stronger centrally-peaked radiation field.
\\

\subsection{Varying bulge stellar population}
\label{subsec:bulgevar}

\begin{figure}
\centering
\includegraphics[width= 3.5in]{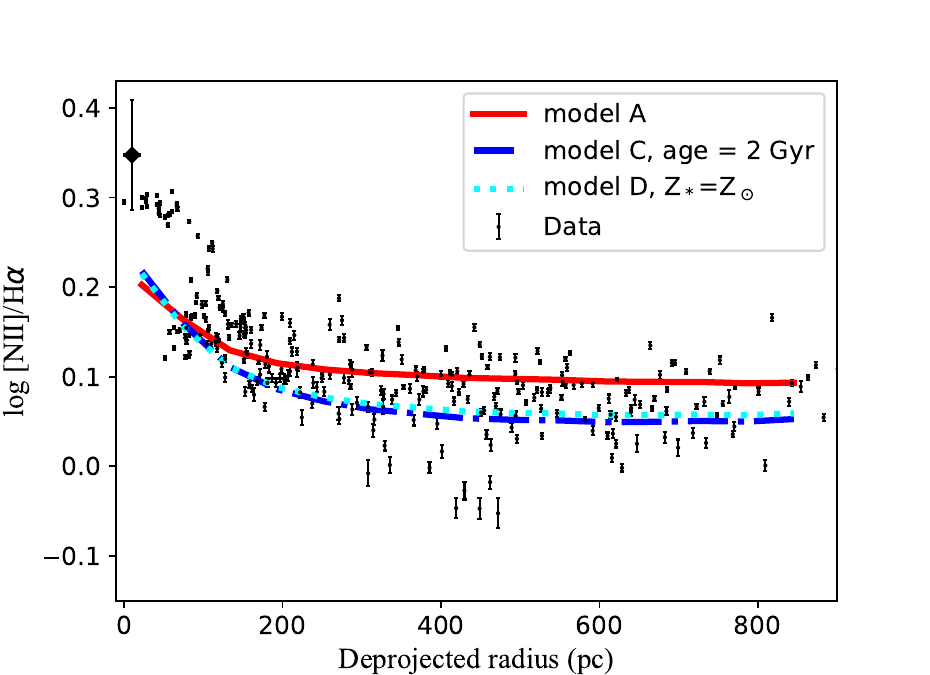}
\includegraphics[width= 3.5in]{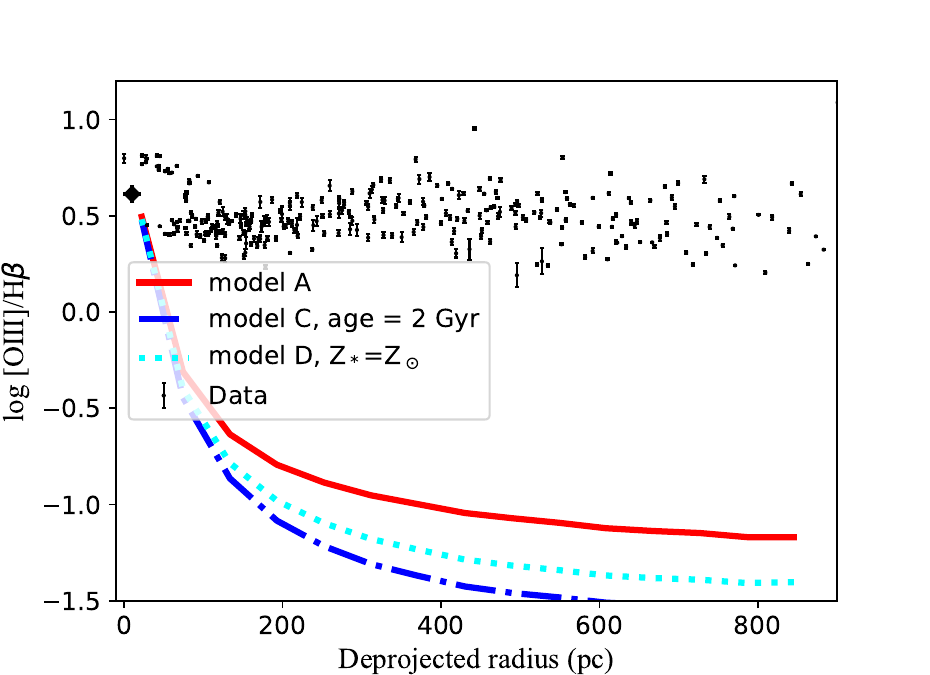}
\caption{
Similar to Figure \ref{fig:high_AGN}, but 
comparing models with various SP properties, as described in the text. Model A with a default bulge SP (age = 8 Gyr, metallicity $Z = 1.5~Z_{\odot}$) is shown as a solid red line. Model D with a younger, 2-Gyr-old SP is shown as a dash-dotted blue line. Model E with a SP with an age of 8 Gyr a solar metallicity SP is represented by a dotted cyan line. The same AGN SED is adopted in these three models.
\label{fig:ratio_deproj}}
\end{figure}

There are two parameters in the SP models that may have a significant influence on the input radiation field and thus on the simulated line intensities: age and metallicity. Therefore, we vary both parameters to examine their effects. 
Model D employs a younger, 2-Gyr-old bulge SP; Model E assumes the same 8 Gyr-old SP but with a solar metallicity. 
The other parameters and conditions remain unchanged as in model A.
The resultant line ratios from these three models, as well as the observed line ratios, are shown in Figure \ref{fig:ratio_deproj}.
\\

It is evident that, just like the default model, both models with an updated SP can more-or-less reproduce the observed trend of the [N\,{\sc ii}]/H$\alpha$ ratio at radii $\gtrsim$ 100 pc. 
The difference between the two models is mild, but their [N\,{\sc ii}]/H$\alpha$ ratios are both systematically lower than the prediction of model A. 
This is presumably due to the ionizing flux of the default bulge SP being somewhat higher than that of the two updated SPs (Figure \ref{fig:sed}).
In terms of the [O\,{\sc iii}]/H$\beta$ ratio, the two models with an updated SP also have a similar behavior as model A, exhibiting a monotonically decreasing profile that is clearly inconsistent with the observed line ratio distribution. 
This is mainly due to the low [O\,{\sc iii}] intensity predicted by the models, as already pointed out in Section \ref{subsec:default}. 
The discrepancy in the two new models is even worse, again presumably due to a lower ionizing flux.
\\

\subsection{Varying nebular properties}
\label{subsec:nebulavar}

\begin{figure}
\centering
\includegraphics[width=3.5 in]{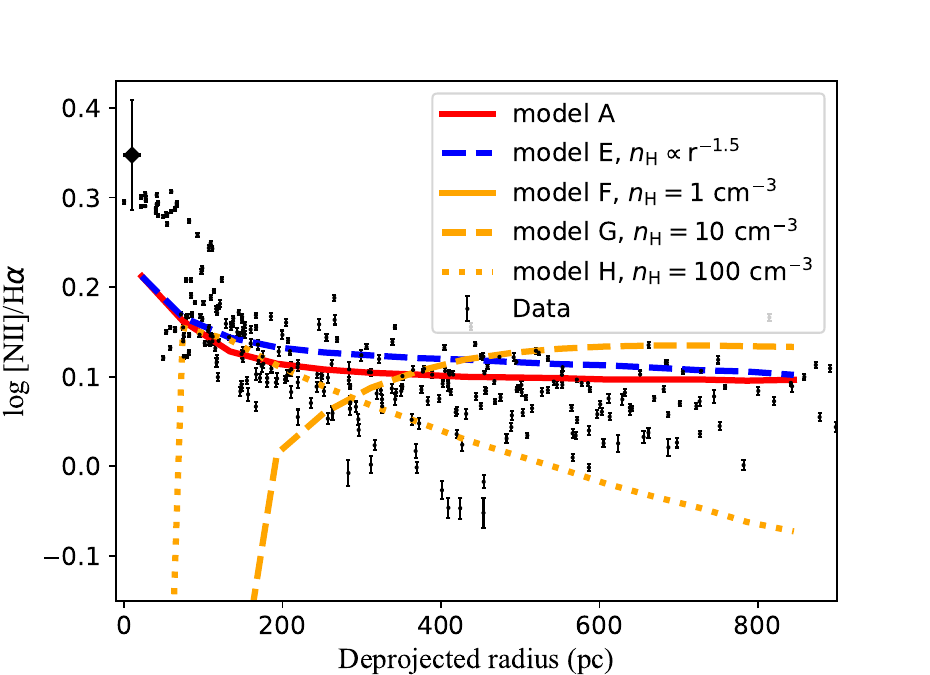}
\includegraphics[width=3.5 in]{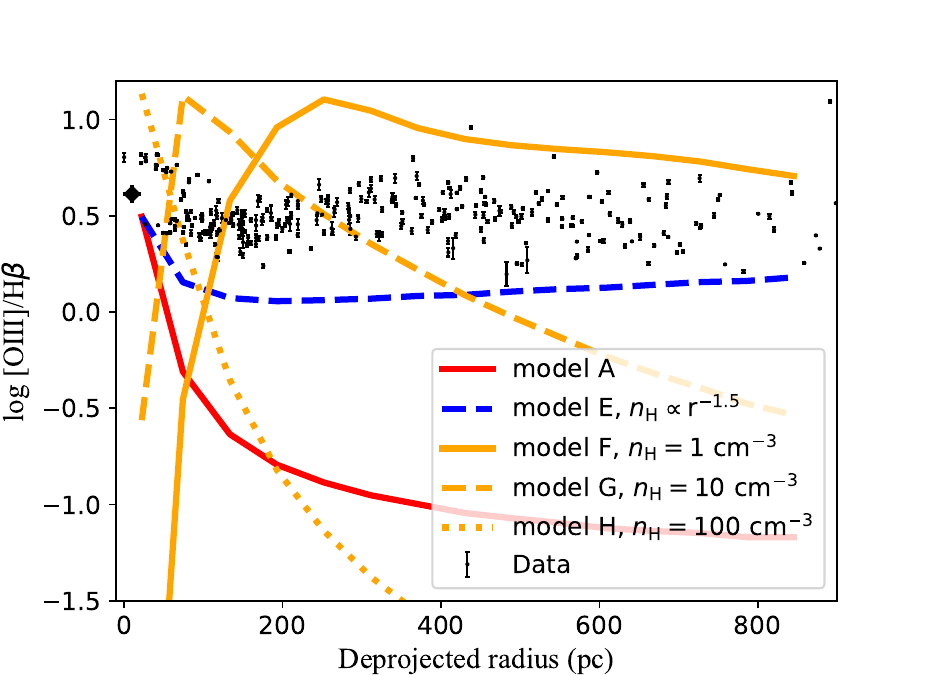}
\caption{$Left$: 
Similar to Figure \ref{fig:high_AGN}, but 
comparing models with different density distributions. The default model A with a density profile $n_{\rm H} \propto r^{-1}$ is shown with a solid red line. Model F with a density profile of $n\rm _H \propto r^{-1.5}$ is shown with a dashed blue line, and models G, H and I with a constant density of 1, 10 and 100 cm$^{-3}$ are shown with a dotted, dashed and solid orange line, respectively. In the left panel, the line ratio predicted by model G is below the range presented here, due to a significantly lower [N\,{\sc ii}] resulted in this model.
\label{fig:ratio_dens}}
\end{figure}

In the models discussed so far, a radial distribution of the hydrogen density $n\rm_H \propto \it r\rm^{-1}$ is adopted, which is roughly consistent with the electron density profile inferred from the [S\,{\sc ii}] doublet line ratios (Figure \ref{fig:ne}). 
Here we assess the effect of varying the hydrogen density profile, by considering four more models: model F assumes $n_{\rm H} \propto r^{-1.5}$ (the dash-dotted line in Figure \ref{fig:ne}), model G assumes a constant $n_{\rm H}$ = 1 cm$^{-3}$, model H assumes a constant $n_{\rm H} = 10 $ cm$^{-3}$, and model I assumes a constant $n_{\rm H} = 100 $ cm$^{-3}$. 
These values bracket the range of observed electron density at a radius $\gtrsim 100$ pc and are representative of the more diffuse part of the circumnuclear ionized gas. 
Since the ionization parameter is directly dependent on the hydrogen density, one expects that varying the density distribution has a significant influence on the emission lines.
The [N\,{\sc ii}]/H$\alpha$ and [O\,{\sc iii}]/H$\beta$ ratios of these three models are shown in Figure \ref{fig:ratio_dens}, again in comparison with model A and the data.
Model F, which follows a steeper density distribution than the default model, produces a similar but slightly higher [N\,{\sc ii}]/H$\alpha$ profile and thus is more-or-less consistent with the data. 
The predicted H$\alpha$, H$\beta$, and [NII] fluxes are in fact similar in both models (Figure \ref{fig:flux_deproj}).
On the other hand, model F predicts a systematically higher and flatter [O\,{\sc iii}]/H$\beta$ profile than the default model, coming much closer to the data.
This can be understood as an inverted ionization parameter profile, which first declines within $\sim$150 pc and then increases outwards, in response to the faster decline of the hydrogen density. More discussions about model F are deferred to Section \ref{sec:5}.
The constant-density models G, H and I show a strong deviation from the observed [N\,{\sc ii}]/H$\alpha$ and [O\,{\sc iii}]/H$\beta$ profiles, although each model overlaps with the observed profile at some radius. 
Here, a constant density combined with a certain ionization parameter at a given radius just acts like the conventional single-zone photoionization model,
which is often able to reproduce a certain line ratio averaged over some physical region, but is unable to track the spatial variation in both parameters in reality. 
This underscores the advantage of our spatially-resolving approach. 
\\

\begin{figure}
\centering
\includegraphics[width=3.5 in]{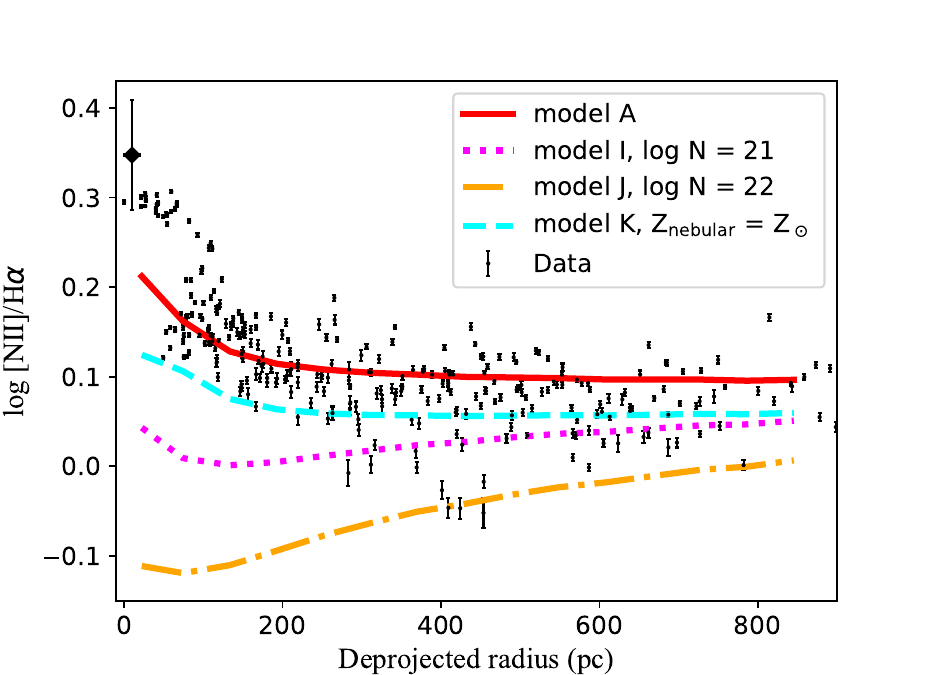}
\includegraphics[width=3.5 in]{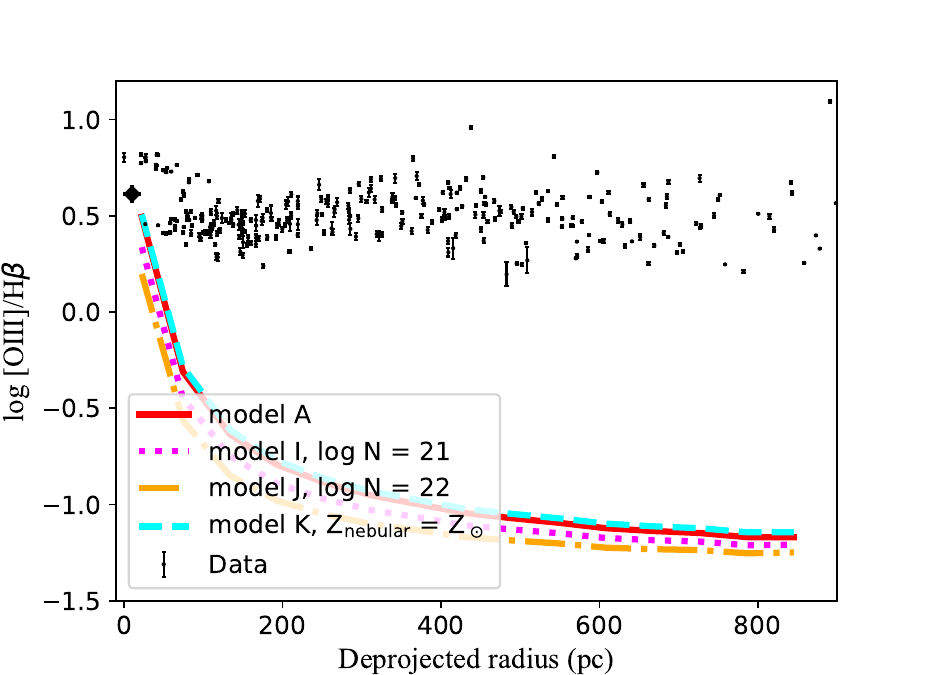}
\caption{$Left$: 
Similar to Figure \ref{fig:high_AGN}, but 
comparing models with  different cloud column densities. The default model A with a column density $N\rm_H = 10^{20}$ cm$^{-2}$ is shown with a solid red line. Model J with $N\rm_H = 10^{21}$ cm$^{-2}$ is shown with a dotted magenta line, while model K with $N\rm_H = 10^{22}$ cm$^{-2}$ is shown with a dash-dotted orange line. Model L with a solar-metallicity gas is shown with a dashed cyan line.
\label{fig:ratio_neb}}
\end{figure}

Besides the density, the cloud column density may also affect the modeled line intensities. Thus, we consider two more models with different column densities: model J takes $N\rm_H = 10^{21}$ cm$^{-2}$, and model K takes $N\rm_H = 10^{22}$ cm$^{-2}$, while the other parameters remain the same. As demonstrated in Figure \ref{fig:ratio_neb}, models J and K result in lower [N\,{\sc ii}]/H$\alpha$ and [O\,{\sc iii}]/H$\beta$ ratios than the default model, thus are inconsistent with the data. 
At last, we test a model with solar metallicity clouds (model L), as is also illustrated in Figure \ref{fig:ratio_neb}, which exhibits a significantly lower ratio in [N\,{\sc ii}]/H$\alpha$ and a slightly lower [O\,{\sc iii}]/H$\beta$, compared to the default model, as a natural consequence of a reduced metallicity. 
We remind that molecules and dust grains are not formally included in our models, but our test simulations taking them into account suggest that the [N\,{\sc ii}]/H$\alpha$ ratio increases by less than 0.1 dex than in the dust-free default model. The difference in the [O\,{\sc iii}]/H$\beta$ ratio is even smaller. The elevation of the predicted line ratios primarily stems from a decrease in the Balmer line intensities, as the cooling from dust grains and formation of molecular clouds cause a mild suppression of the ionized zone.
\\

\subsection{$\chi^2$ estimation of different models}

To be more quantitative about the difference between the observed data and various models, we calculate for each model the $\chi^2$ following the definition of Pearson's chi-square test: 
\begin{equation}
    \chi^2 = \frac{1}{N}\Sigma\frac{(I\rm_{obs}-\it I\rm_{model})^2}{\it I\rm_{model}^2}
\end{equation}
where $I\rm_{obs}$ represents the observed value at each Voronoi bin, $I\rm_{model}$ the model-predicted value interpolated at each bin, and $N$ the number of spatial bins. 
Here we evaluate the $\chi^2$ values for the line intensities of H$\alpha$ and [O\,{\sc iii}] and line ratios of [N\,{\sc ii}]/H$\alpha$ and [O\,{\sc iii}]/H$\beta$, as representative results. 
It is important to note that our model comparison is not taken as a formal fitting to the data.  
Moreover, the exact number of free parameters in the {\sc cloudy} models cannot be precisely counted, although it should be the same for all models. 
Therefore, the $\chi^2$ presented here are meaningful only in a relative sense. 
\\

\begin{figure}
\centering
\includegraphics[width= 3.5in]{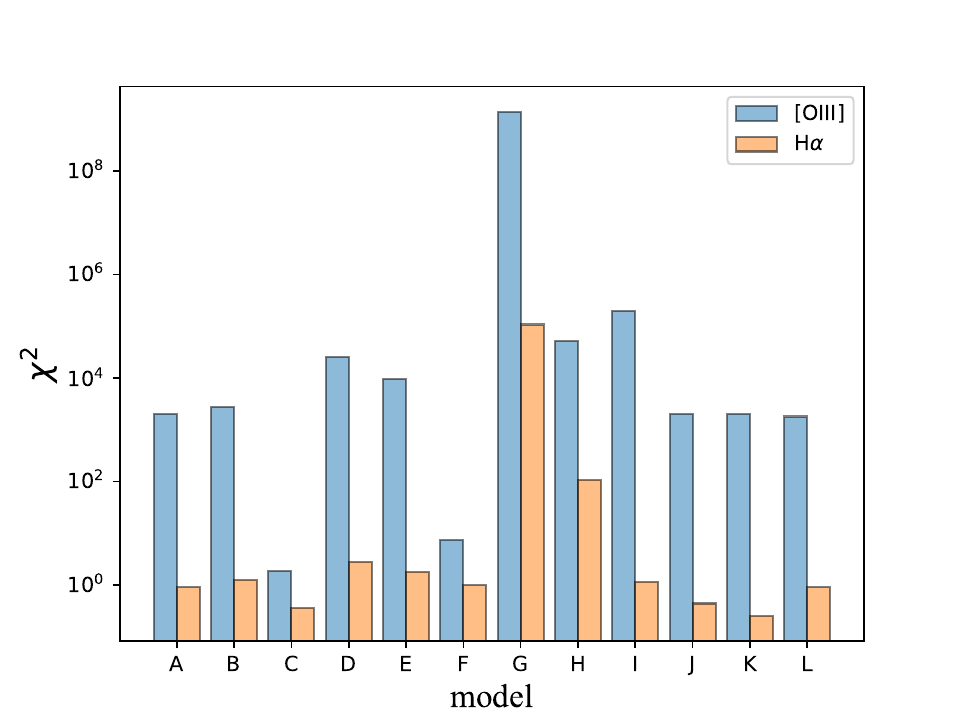}
\includegraphics[width= 3.5in]{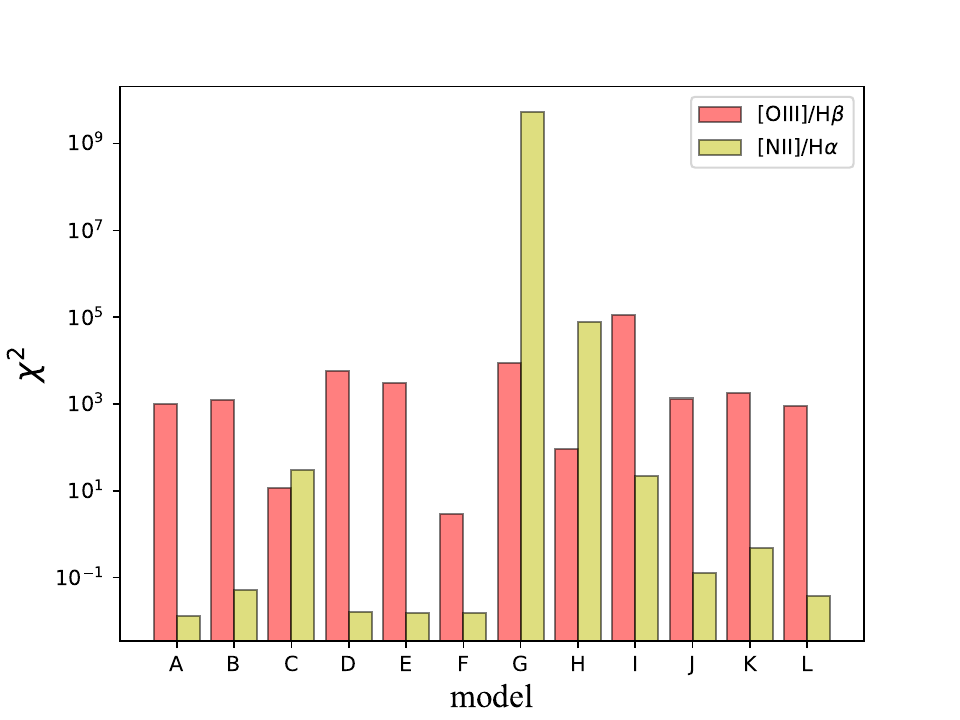}
\caption{The $\chi^2$ distribution of different models, for the line intensities ({\it left}) and line ratios ({\it right}).
\label{fig:chi2}}
\end{figure}

The resultant $\chi^2$ are presented in Figure \ref{fig:chi2}. 
Regarding H$\alpha$, it is evident that the $\chi^2$ values are generally smaller compared to [O\,{\sc iii}]. With the exception of model G ($n = 1 ~\rm cm^{-3}$) and model H ($n = 10 ~\rm cm^{-3}$), all models exhibit relatively small $\chi^2$ values for H$\alpha$. 
However, for the [O\,{\sc iii}] line, the $\chi^2$ values are large for all models except model C (AGN*10+SP) and model F ($n\propto r^{-1.5}$). As for the line ratios, models A, B, D, E and F yields similarly small $\chi^2$ values for [N\,{\sc ii}]/H$\alpha$. On the other hand, only models C and F show a relatively small $\chi^2$ values for the [O\,{\sc iii}]/H$\beta$ ratio. 
The behavior of $\chi2$ in Figure \ref{fig:chi2} is basically consistent with our earlier assessments about the individual models. 
Apparently, the gas density profile and AGN luminosity are two promising parameters for alleviating the deficiency of the model-predicted [O\,{\sc iii}]. The effects of these parameters will be further discussed below.
\\

\section{Discussion}
\label{sec:5}
In previous sections, we have investigated the scenario of photoionization for one of the nearest LINERs, the one situated in the massive spiral galaxy M81, which is well resolved by our IFS observation. 
We confront the radial distribution of major emission lines, in particular H$\alpha$, H$\beta$, [N\,{\sc ii}] and [O\,{\sc iii}], with a set of {\sc cloudy} photoionization models. 
A novel aspect of these models is that their main input parameters are empirically well constrained, reflecting the well-established properties of a LLAGN and an old, massive stellar bulge.  
The four lines, frequently used as diagnostics of the warm ionized gas, behave remarkably similar in the central few hundred parsecs, in the sense that their intensities all decline by about two orders of magnitude from a deprojected radius of $\sim$20 pc to $\sim$800 pc (Figure \ref{fig:eml_agn_sp}). 
Accordingly, the [N\,{\sc ii}]/H$\alpha$ and [O\,{\sc iii}]/H$\beta$ ratios show a similar radial trend as well, both declining within $\sim$150 pc and remaining flat outwards (Figure \ref{fig:ratio_agn_sp}). 
Our models, which combine ionizing photons from a central LLAGN and a spatially extended old stellar population, have only a limited degree of success in matching the observation.
In particular, it is found that some of our models (especially the default model) can reproduce the observed radial trend of H$\alpha$, H$\beta$ and [N\,{\sc ii}] reasonably well, although the model-predicted absolute line intensities fall short by $\sim$50\% on average. 
On the other hand, essentially all models fail to reproduce the observed [O\,{\sc iii}] line distribution, as the model-predicted [O\,{\sc iii}] profile almost always declines too rapidly. 
Meanwhile, some models can reproduce the [N\,{\sc ii}]/H$\alpha$ ratio, whereas all models exhibit a large discrepancy with the observed flat distribution of the [O\,{\sc iii}]/H$\beta$ ratio at large radii. 
This can be understood as a deficiency of ionizing photons in the outer region, despite the presence of a spatially extended stellar bulge that fully encompasses the LINER. 
\\

As described in Section~\ref{sec:3}, we have tried to constrain our model parameters and settings from existing observations as much as possible. Some parameters, such as the SEDs of the LLAGN and the SP, are well-motivated and strongly constrained. 
However, some other parameters, in particular those related to the clouds, are less well determined, but they could have a significant influence on the results (Section~\ref{subsec:nebulavar}), which may affect our general conclusion about the success (or failure) of the photoionization scenario. 
Below we address some relevant issues. 
\\

\subsection{Geometric effect}
We first consider the potential influence of geometric effect on our modeling. 
As illustrated by Figure \ref{fig:geo}, a basic assumption in our models is that the illuminated clouds are coplanar with the galactic gas/stellar disk, which is physically plausible if much of the circumnuclear gas, especially those composing the nuclear spiral, is fed by the large-scale disk.
Still, this assumption could be relaxed, as the inclination angles of the nuclear spiral and the galactic disk are not necessarily the same, as suggested for the case of M31 \citep{1988AJ.....95..438C}.
As a test, we make the alternative assumption that the spatial radius of the Voronoi bins exactly equals their projected distance from the galactic center. 
This is illustrated in Figure \ref{fig:flux_deproj}, which is similar to Figure \ref{fig:eml_agn_sp} but adding a set of green data points, which have acquired their radial position by shifting the black data points inward, i.e., without deprojection.
In this case, a significantly better agreement is achieved between the observed and modeled line intensity profiles of H$\alpha$, H$\beta$, and [N\,{\sc ii}], not only for the general shape but also for the absolute strength. In fact, we have tested and found that the $\chi^2$ values of the undeprojected case is significantly smaller than the deprojected case for all emission lines.
However, shifting the clouds to the new locations is still far from obtaining a satisfactory modeling for the oxygen lines, especially [O\,{\sc iii}]. 
\\

A further complication is that
some fraction of the ionized gas is distributed out of the assumed plane and spreading within the bulge, which, for example, could be related to a bipolar outflow driven by the LLAGN or filaments cooling from a pre-existed hot gas corona.
In this case, the observed line intensity would be an integration of all ionized gas along the line-of-sight, which includes contributions from the bulge component. 
As mentioned in Section \ref{subsec:other}, the incident radiation field of out-of-plane clouds is 0.2--0.5 dex lower than that of clouds in the disk within a radius of 200 pc, while being nearly the same at larger radii, an effect due to the extended stellar distribution. 
Hence, including the potential contribution from a low-density, diffuse ionized gas (DIG) in the bulge could enhance the predicted line intensities, especially for [O\,{\sc iii}] at outer radii, thus alleviating the discrepancy between the observation and models. The potential contribution of a bulge DIG can be approximated with model G presented in Figure \ref{fig:ratio_dens}, which adopts a constant density of 1 cm$^{-3}$. It is evident that this model primarily contributes to the [O\,{\sc iii}] in the outer region, while exerting negligible influence on the other lines.  
However, the very existence of a bulge DIG in M81 is not yet established. Moreover, it would probably require some fine-tuning of the DIG properties to raise the [O\,{\sc iii}] intensity by a factor of 10--100 while only a factor of two or less for the Balmer lines (Figure \ref{fig:flux_deproj}), so as not to substantially violate the observation. 
In this case, the Balmer lines at the outer regions would still be dominated by the default disk, while the [O\,{\sc iii}] line arises mainly from the bulge DIG, to some extent similar to the case of extraplanar DIG seen in star-forming galaxies \citep{2022A&A...659A..26B}.
However, we consider such a case highly unlikely because no significant difference in the velocity field of H$\alpha$ and [O\,{\sc iii}] is observed \citepalias{2022ApJ...928..111L}, which is otherwise expected for two spatially distinct components. 
Future IFU observations with a high spectral resolution (better than few tens of km~s$^{-1}$) would be crucial to isolate the bulge ionized gas, if it truly exists. 
\\

\subsection{Density distribution}
\begin{figure}
\centering
\includegraphics[width= 3.5in]{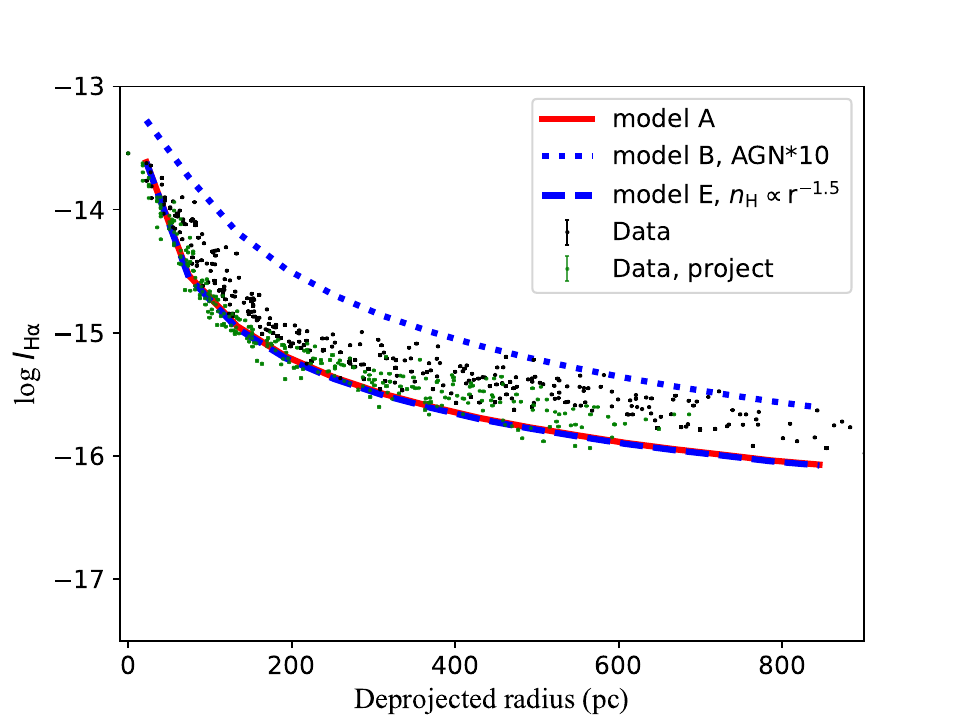}
\includegraphics[width= 3.5in]{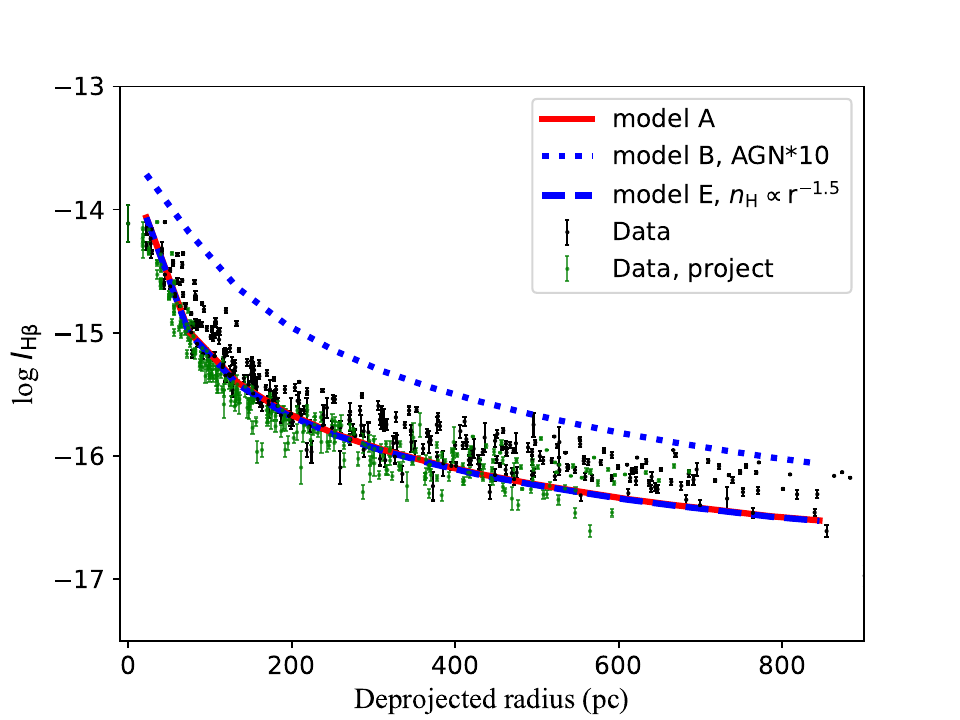}
\includegraphics[width= 3.5in]{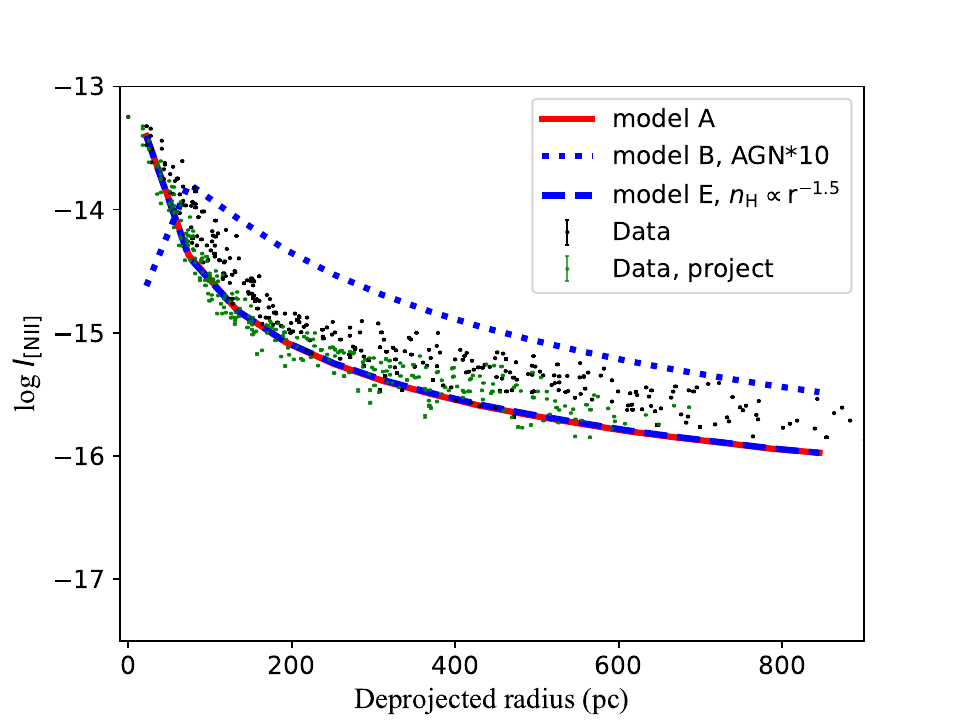}
\includegraphics[width= 3.5in]{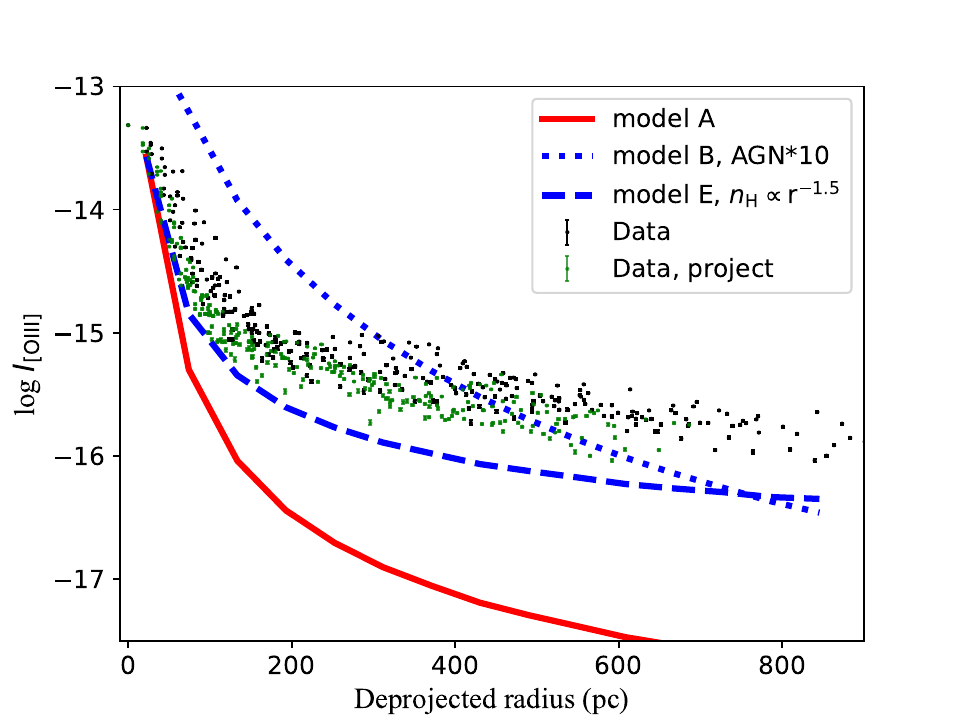}
\includegraphics[width= 3.5in]{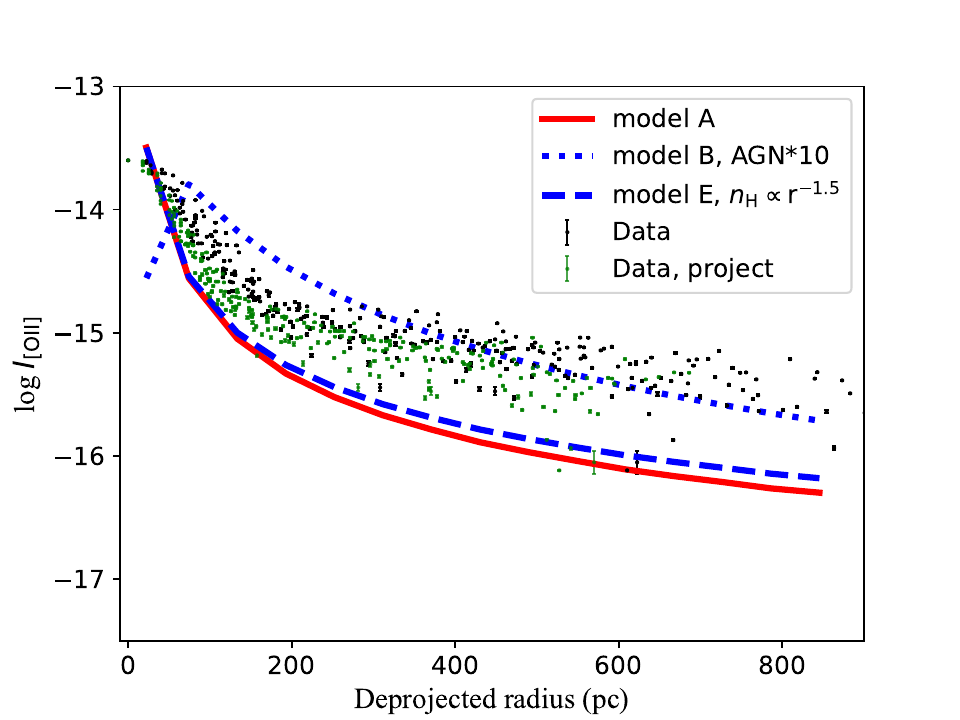}
\includegraphics[width= 3.5in]{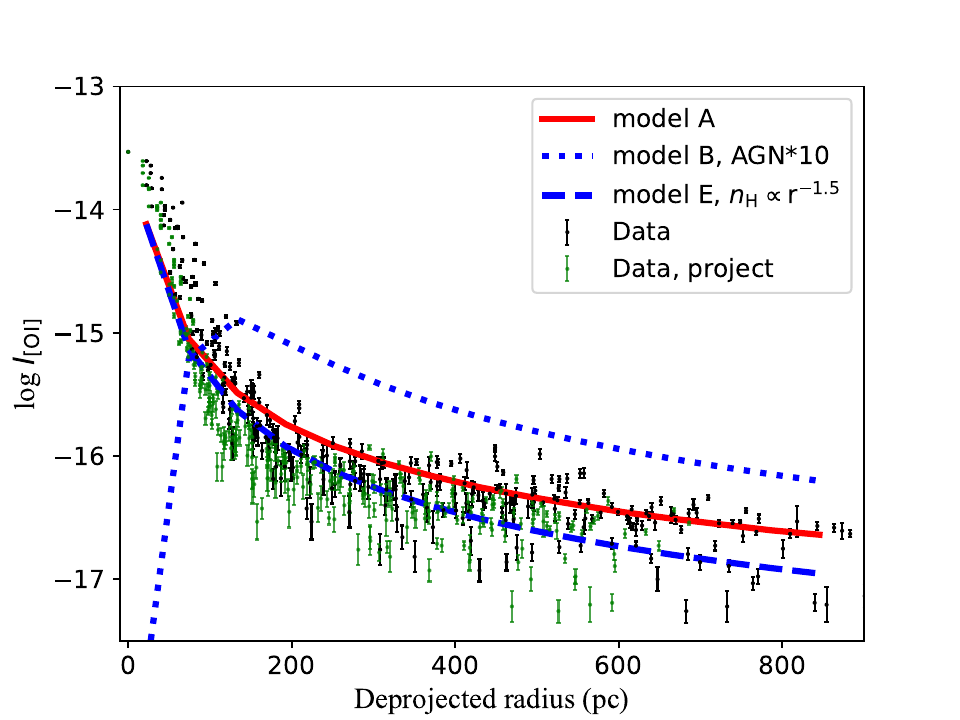}
\caption{The radial distribution of the observed line intensities and the simulated line intensities from the {\sc cloudy} models of the major emission lines: H$\alpha$, H$\beta$, [N\,{\sc ii}], [O\,{\sc iii}], [O\,{\sc ii}] and [O\,{\sc i}]. The data points in black are the same as in Figure \ref{fig:eml_agn_sp}. 
The data points in green show the observed line intensity distribution without deprojection, that is, assuming an inclination angle of zero. 
The solid red lines represent the default model A, the dotted blue  line represents model C with a 10 times brighter AGN, and the dashed blue line represents model F with a density profile of $n{\rm_H} \propto r^{-1.5}$. The latter two models have a better chance to alleviate the discrepancy between the observed flat trend of [O\,{\sc iii}]/H$\beta$ ratio and the prediction of the default model. 
\label{fig:flux_deproj}}
\end{figure}

\begin{figure}
\centering
\includegraphics[width= 3.5in]{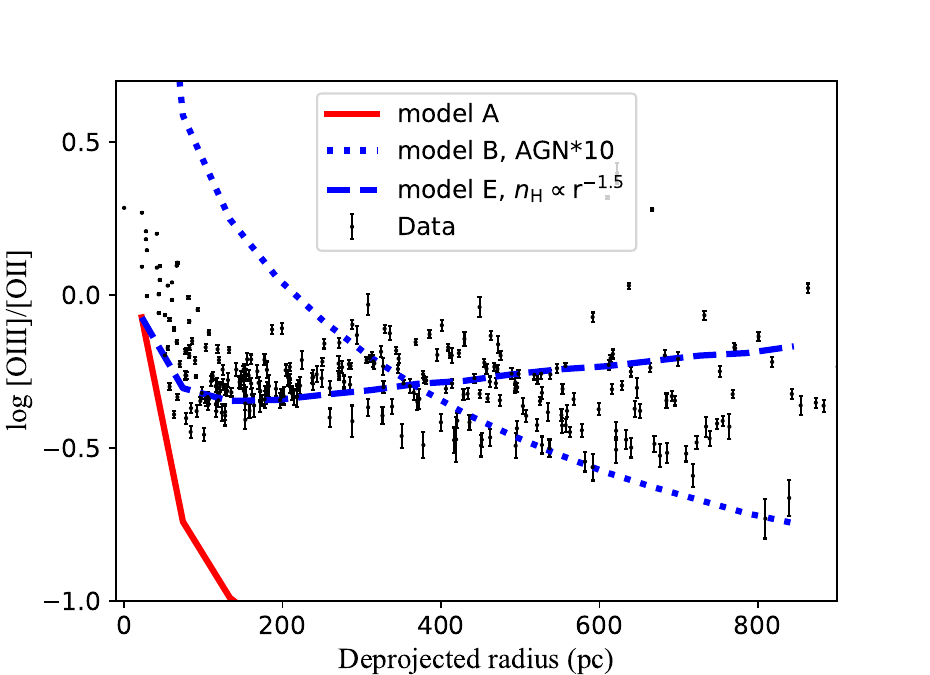}
\includegraphics[width= 3.5in]{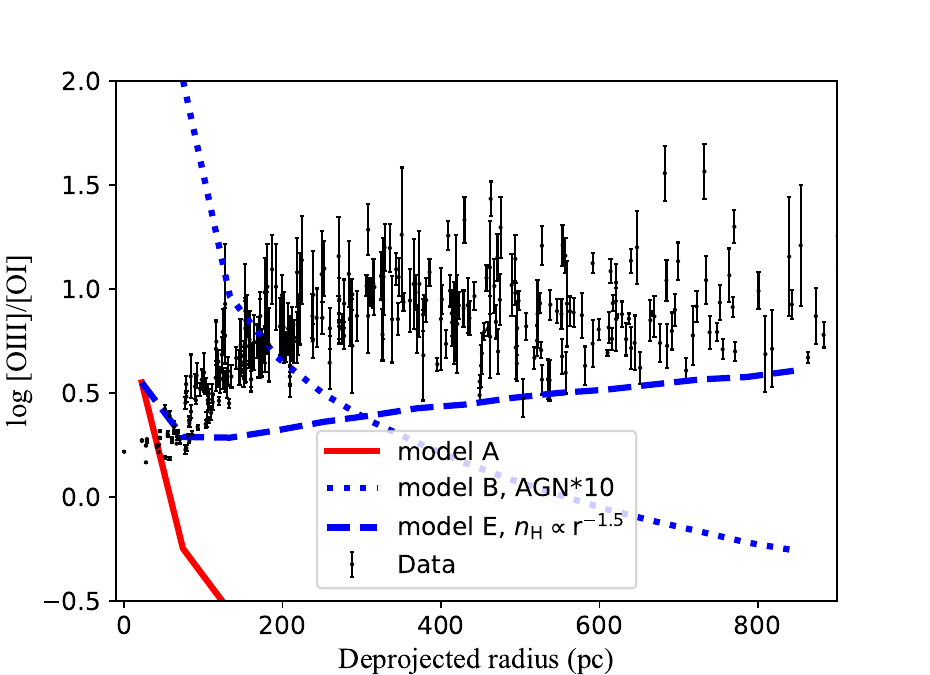}
\caption{The radial distribution of [O\,{\sc iii}]/[O\,{\sc ii}] ({\it left}) and [O\,{\sc iii}]/[O\,{\sc i}] ({\it right}) ratios. Symbols are the same as in Figure \ref{fig:flux_deproj}.
\label{fig:OIII_OII}}
\end{figure}

Another parameter that has a significant effect on the modeled line intensity is the 
radial distribution of the cloud density, which basically controls the radial variation of the ionization parameter. 
Our default choice of $n{\rm_H} \propto r^{-1}$ was mainly driven by the observed [S\,{\sc ii}] line ratio. However, given the substantial scatter and uncertainty in the [S\,{\sc ii}] line ratio, a steeper density distribution, $n{\rm_H} \propto r^{-1.5}$, is still roughly compatible with the data (Figure \ref{fig:ne}). 
The resultant line intensity profiles of model F, which incorporates this steeper density distribution, are depicted in Figure \ref{fig:flux_deproj}.
Compared to the default model, the match with the observed H$\alpha$, H$\beta$, and [N\,{\sc ii}] profiles stay satisfactory. More importantly, this model, with a more slowly declining ionization parameter, does a much better job of matching the oxygen lines. 
In particular, the discrepancy with the [O\,{\sc iii}] line reduces to within about half dex. 
This suggests that an even steeper (e.g., $n{\rm_H} \propto r^{-2}$) density distribution might completely remove the discrepancy. 
Nevertheless, we consider such a steep distribution of the circumnuclear gas highly unlikely, which would be hardly compatible with the electron density distribution inferred from the [S\,{\sc ii}] line ratio (Figure \ref{fig:ne}).
In principle, only a bulk outflow with a roughly constant velocity is conceivable to acquire a density profile following $r^{-2}$, but such an outflow across the central kpc of M81 is certainly not observed \citepalias{2022ApJ...928..111L}. 
Moreover, the change of the predicted [O\,{\sc iii}] profile comes with changes in the [O\,{\sc ii}] and [O\,{\sc i}] profiles. 
This is illustrated in Figure \ref{fig:OIII_OII}, in which the observed [O\,{\sc iii}]5007/[O\,{\sc ii}]3727 and [O\,{\sc iii}]5007/[O\,{\sc i}]6300 profiles are plotted against models A, C, and F. 
These two line ratios are a direct proxy of the ionization state of oxygen. 
The observed [O\,{\sc iii}]/[O\,{\sc ii}] ratio decreases rapidly within the central 100 pc and then keeps nearly constant outwards, while the [O\,{\sc iii}]/[O\,{\sc i}] ratio increases gradually within 200 pc, and then keeps constant outwards. 
The failure of model A is clearly underscored by this figure.
Even model F can only roughly match the observed [O\,{\sc iii}]/[O\,{\sc ii}] profile, but not [O\,{\sc iii}]/[O\,{\sc i}]. 
Raising [O\,{\sc iii}] with a steep gas density (thus a higher ionization parameter) would inevitably simultaneously decrease [O\,{\sc ii}] and [O\,{\sc i}].
\\

Therefore, it is unlikely that a fine-tuned gas density distribution would provide a satisfactory fit to all three oxygen lines at large radii.
Nevertheless, it is still imperative for future observations to obtain a more accurate measurement of the electron density in order to better constrain the ionization parameter at large radii.
In principle, a potentially better approach is to estimate the electron density using infrared line ratios, such as [O\,{\sc iii}]$\lambda \rm 88 \mu m$/[O\,{\sc iii}]$\lambda \rm 52 \mu m$ and [N\,{\sc ii}]$\lambda \rm 205\mu m$/[N\,{\sc ii}]$\lambda \rm 122 \mu m$, which are more sensitive to lower electron densities \citep{2019ARA&A..57..511K}.
\\

\subsection{Previous AGN activity?}
Changing the AGN bolometric luminosity, the other key factor controlling the ionization parameter also does not provide a satisfactory fit to the data. This is again highlighted in Figure~\ref{fig:flux_deproj}, showing model C which adopts a 10 times higher AGN bolometric luminosity.
Compared to the default model, the discrepancy with the observed line profiles becomes worse except perhaps for [O\,{\sc iii}] and [O\,{\sc ii}]. 
Even for these two lines, the model-predicted profiles exhibit a clear imbalance between the inner and outer regions, resulting in a mismatch to the observed [O\,{\sc iii}]/[O\,{\sc ii}] and [O\,{\sc iii}]/[O\,{\sc i}] profiles (Figure \ref{fig:OIII_OII}).
Model C generally predicts too high line intensities for the other four lines, except for a lower [N\,{\sc ii}] and [O\,{\sc i}] within $\sim$100 pc.
\\

While a constantly higher AGN luminosity fails to reproduce the observation, a more active M81* in the recent past, which then quenched and resumed to the low level as observed now, seems to be viable for alleviating the deficiency of ionizing photons for the outer regions, as consistently hinted by our models. 
A fading AGN has been proposed for a number of nearby galaxies harboring extended (up to 10 kpc) emission-line regions showing Seyfert characteristics \citep{2009MNRAS.399..129L, 2022ApJ...936...88F, 2022ApJ...938...75P}. Since highly ionized species recombine more quickly in the central regions where the electron density is generally higher, when taking the light-crossing time into account, the strong [O\,{\sc iii}] emission in the outer regions might be the relic of previous AGN activities. 
\\

More quantitatively, the recombination timescale of a certain ion follows $\tau \sim 1/({n\rm_e \alpha\rm_A})$,
where $\alpha\rm_A$ is the recombination coefficient, which can be obtained from pyatomdb \citep{2020Atoms...8...49F} based on the AtomDB database\footnote{https://www.atomdb.org}. 
The recombination coefficient of O$^{++}$ is $\rm \sim 3 \times 10^{-12}~cm^3~s^{-1}$ at $T\rm_e = 10^4~K$ under Case B recombination. In the central 100 pc, where the electron density is $\rm \gtrsim 10^2~cm^{-3}$, we have $\tau(\rm O^{++}) \sim$100 years, while in the outer regions where $n\rm_e \sim 10~cm^{-3}$, it takes ten times longer, $\sim 1000$ years, for O$^{++}$ to become O$^{+}$. It is noteworthy that O$^{++}$ might be replenished through the recombining higher state ions, such as O$^{3+}$. However, the recombination rate of these higher states is considerably greater, resulting in a shorter recombination timescale and thus a negligible supply of O$^{++}$.  In comparison, the recombination time scale for hydrogen is approximately one order of magnitude longer for the same electron density. 
Therefore, if M81* had for some reason raised its luminosity substantially several thousand years ago for only a relatively short period, the outward propagating ionizing light front would produce both highly and lowly ionized ions from inside out.
Then, as M81* quickly faded,
most doubly-ionized oxygen in the central 100 pc would  recombine to O$^+$, resulting in a [O\,{\sc iii}]/H$\beta$ ratio not significantly different from the value dedicated by the presently low-state of M81*.
In contrast, in the outer regions the recombination of O$^{++}$ takes longer, thus the [O\,{\sc iii}]/H$\beta$ ratio remains high. 
As for [N\,{\sc ii}], because the recombination coefficient of N$^{++}$ is approximately twice higher than O$^{++}$, 
a recovery of the [N\,{\sc ii}]/H$\alpha$ ratio in the central region should occur within an even shorter period. Thus, the flat distribution of both line ratios seems to be qualitatively explained under such a scenario. 
Nevertheless, comparing models A and C in Figure \ref{fig:flux_deproj}, the intensity of the Balmer lines roughly scales with AGN luminosity. 
This means that for a flaring AGN with only a 10-times boosted luminosity, the expected H$\alpha$ and H$\beta$ intensities at large radii, where hydrogen recombines on a timescale of $\sim 10^{4}$ year, may still be considered in rough consistency with the observation,  
But more energetic AGN eruptions within $\sim 10^{4}$ year can be ruled out given the present Balmer line intensities.
A more quantitative study of the faded AGN scenario will be reserved for future work. 
\\

\subsection{Stellar ionizing sources}

We have implicitly assumed that the hot low-mass evolved stars are expected to account for the stellar photoionizaton. 
Among the HOLMES populations, 
the majority of ionizing photons are thought to be  provided by pAGB stars \citep{2011MNRAS.415.2182F}. This is support by our SSP models compiled from FSPS.
To illustrate this, 
we show in the left panel of Figure \ref{fig:sed} the SED (cyan dash-dotted line) of a sum of 10$^4$ pAGB stars, each with a blackbody temperature of 10$^5$ K and a luminosity of 10$^3~L_\odot$. 
We have assumed that the synthetic pAGB stars follow the same deprojected S$\rm \acute{e}$rsic distribution as the bulge. It can be seen that they can roughly reproduce the UV flux of the total bulge SP at a galactocentric radius of $\sim$20 pc. Notably, the abundance of these synthetic pAGB stars, $4\times10^{-7}\rm~M_\odot^{-1}$, is in rough agreement with the abundance ($\sim 2\times10^{-7}\rm~M_\odot^{-1}$) of UV-bright stars (mainly pAGB stars and post-early AGB stars) detected in the inner bulge of M31 by the Panchromatic Hubble Andromeda Treasury program \citep{2012ApJ...755..131R}, which is presently the only massive stellar bulge where such a UV population can be directly resolved. 
This indicates that the pAGB stars are indeed the dominant source of stellar ionizing photons. 
\\

We note that the UV ionizing flux of one HOLMES is roughly equivalent to the sum of $10^6$ WDs with a similar blackbody temperature of $10^5$ K but a typical luminosity of only 10$^{-3}~L_\odot$ \citep{2005ApJS..156...47L}. In the Milky Way, the total number of WDs is estimated through the extrapolation of local WDs, which is at most $10^{10}$ \citep{2009JPhCS.172a2004N}, including a large population in the halo. 
Given the fact that M81 has a similar total stellar mass as the Milky Way, its WD population should also be similar in number and thus must fall short to provide an ionizing flux equivalent to that by the HOLMESs. 
Another possible ionizing source, not included in our SP model, is the PNe. However, PNe are an insignificant contributor to the ionizing energy budget, as is indicated in previous studies. Following \cite{2012ApJ...747...61Y}, the [O\,{\sc iii}] luminosity produced by PNe is 1.35 $\times 10^{28} ~ L\rm_{gal}/\it L\rm_\odot~erg~s^{-1}$ adopting a PN abundance of 1.65 $\times 10^{-7} ~ L\rm_{gal}/\it L\rm_\odot$ from a sample of early-type galaxies \citep{2006MNRAS.368..877B}. Therefore, assuming
a total stellar mass of $\sim 2 \times 10^{10}~M_\odot$, this is equivalent to a total [O\,{\sc iii}] luminosity of $\sim 3\times 10^{38}$ erg s$^{-1}$. This value is much lower than the observed [O\,{\sc iii}] luminosity within the FoV of $2.2 \times 10^{39}$ erg s$^{-1}$. Hence, we can conclude that PNe cannot account for the observed [O\,{\sc iii}] line emission.
\\

Another possible stellar ionizing source is XRBs, in particular low-mass X-ray binaries (LMXBs), which have a much harder SED than the UV-bright stars.  
We assume 200 bulge LMXBs following the deprojected S$\rm \acute{e}$rsic distribution, which have a luminosity function determined from {\it Chandra} X-ray observations of M81, $n({\rm XRB})\propto L^{-1}$ \citep{2010ApJ...721.1523K, 2019IAUS..346..344S}. The SED of a single LMXB is assumed to be a broken power law with a photon index of -1.56, as derived from the LMXBs in 15 nearby early-type galaxies \citep{2003ApJ...587..356I}. The overall SED and the ionization parameter of the XRBs are shown in Figure \ref{fig:sed}. 
Evidently, the LMXBs are at best a minor source of ionization compared to the UV-bright stellar populations. This holds true even if we consider the effect of secondary ionization by photoelectrons, which is $\sim$10 per primary X-ray photon \citep{1979ApJ...234..761S}.
Our test simulations taking into account this LMXB population supports our assessment,
which is also consistent with the argument by \cite{2010MNRAS.402.2187S} that LMXBs would produce much fewer ionizing photons than pAGB stars.
Similarly, diffuse hot gas, which is known to present in the M81 bulge \citep{2003ApJS..144..213S}, is insufficient as an ionizing source due to an even lower X-ray luminosity compared to the LMXBs.
High-mass X-ray binaries (HMXBs) and ultraluminous X-ray sources (ULXs) can be safely neglected in the central kpc region of M81 due to the lack of young stellar populations.
\\

The deficiency of stellar photoionization for the M81 LINER has an important implication for LIERs, i.e., low-ionization emission regions with an extent up to galactic scales \citep{2012ApJ...747...61Y, 2016MNRAS.461.3111B}. 
Flat line ratios with a relatively high value (e.g., log\,[O\,{\sc iii}]/H$\beta \sim 0.4$ and log\,[N\,{\sc iii}]/H$\alpha \sim 0.0$ in a large sample of SDSS red galaxies; \citealp{2012ApJ...747...61Y}) are also observed and taken to be characteristic of LIERs. 
It is suggested that stellar sources, in particular HOLMES, are the most plausible ionization mechanism for LIERs to explain their extended intensity profiles and flat line ratios \citep{2016MNRAS.461.3111B}. 
The photon budget of HOLMES is generally considered sufficient to account for the H$\alpha$ luminosity, although \citet{2012ApJ...747...61Y} already pointed out that the requirement of the ionized gas fully absorbing the stellar ionizing photons may not be fulfilled.
However, the inclusion of extended SPs in our models still fails to explain the flat [O\,{\sc iii}]/H$\beta$ profile, clearly pointing to a deficiency of ionizing photons at outer radii.
Following the common practice, we calculate the total ionizing photon from the HOLMESs using the bulge SP as presented in Section \ref{subsec:default}. The ionizing photon rate produced within the central 600 pc is $Q\rm_H \sim 1.5 \times 10^{51}$ ph~s$^{-1}$. Assuming all these ionizing photons are completely absorbed by the clouds, and it takes 2.2 ionizing photons to produce one H$\alpha$ photon, this will produce a H$\alpha$ luminosity of 2 $\times$ 10$^{39}$ erg s$^{-1}$, which well matches with the observed H$\alpha$ luminosity of $\sim$2 $\times$ 10$^{39}$ erg s$^{-1}$ within the same radius. 
Such a situation is quite similar to the case of LIERs \citep{2016MNRAS.461.3111B}.
However, we have demonstrated that the stellar photonization fails to account for the overall observed line distributions in the central kpc region of M81. It would be interesting to revisit the feasibility of stellar photonization as a prevailing mechanism for the LIERs, taking into account the detailed geometry of the stars and gas as is done for M81 here.
\\

\section{Summary}
\label{sec:6}

In this work, we have conducted a comprehensive investigation of the conventional photoionization scenario for one of the most extensively studied LINERs, which is located in the nearby galaxy M81 hosting a prototypical LLAGN and a massive stellar bulge. 
A set of {\sc cloudy} photoionization models are constructed and confronted with the spatially-resolved emission line properties derived from our CAHA 3.5m IFS observations of the central 1 kpc region of M81. 
The main parameters of these models are well constrained by extensive observations of the LLAGN and the stellar bulge, but also explore a reasonably wide range of uncertain nebular properties, thus allowing for a most detailed comparison between the model-predicted and observed emission lines from the circumnuclear ionized gas. Our main findings are as follows:

\begin{itemize}
    \item Under the reasonable assumption that the bulk of the circumnuclear ionized gas is located within a thin plane defined by the large-scale disk, the combination of a central LLAGN and an extended stellar bulge (AGN+SP) as the photoionization sources can roughly reproduce the observed radial intensity distributions of H$\alpha$, H$\beta$ and [N\,{\sc ii}], but generally have difficulty in producing a similarly declining profile of the [O\,{\sc iii}] line as observed. 
    
    \item The LINER of M81 exhibits a characteristic flat profile in the [N\,{\sc ii}]/H$\alpha$ and [O\,{\sc iii}]/H$\beta$ line ratios out to a deprojected radius of $r \sim$ 800 pc. 
    The AGN+SP model can reproduce the [N\,{\sc ii}]/H$\alpha$ profile well, but declines too fast to reproduce the [O\,{\sc iii}]/H$\beta$ profile, which can be understood as a deficiency of high-energy ionizing photons at $r \gtrsim 200$ pc. 
    This might be reconciled if the bulk of the [O\,{\sc iii}] line arises from a low-density ionized gas spreading across the bulge, but current data provides no compelling evidence for such a case.
    
    \item The gas density profile and AGN luminosity are identified as the most important factors in determining the line ratio distribution, while the SP properties, gas metallicity and column density have a minor impact on the results. 
\end{itemize}

Besides AGN and stellar photoionization, 
a few other ionization mechanisms may also play a role in the gas ionization, potentially responsible for the flat line ratio profile in M81. 
Shocks are a popular explanation for the ionization of gas at large radii \citep{Molina_2018}. The abnormal H$\alpha$/H$\beta$ line ratio (Section~\ref{subsec:default}), if intrinsic, could be a manifestation of shocks \citep{1980A&A....87..152H, 2012MNRAS.419.1402G}, since this ratio can be significantly higher only when collisional heating, such as induced by shocks, dominates the excitation of hydrogen. Cosmic rays are an important source of heating and may boost the intensity of [O\,{\sc iii}] under high cosmic ray densities \citep{1984ApJ...286...42F}. 
A detailed investigation of these candidates is beyond the scope of the present paper and will be reserved for future work. Higher spatial and spectral resolution and multi-wavelength data are required to distinguish the contribution of shocks, isolate various diffuse gas components, and yield better constraints on the gas density distribution. 
\\

\begin{acknowledgments}
This work is supported by the National Key Research and Development Program of China (NO.2022YFF0503402) and National Natural Science Foundation of China (grant 12225302). It is based on observations collected at Centro Astron\'{o}mico Hispano en Andaluc\'{i}a (CAHA) at Calar Alto, operated jointly by Junta de Andaluc\'{i}a and Consejo Superior de Investigaciones Cient\'{i}ficas (IAA-CSIC). 
Z.N.L. acknowledges the fellowship of China National Postdoctoral Program for Innovation Talents (grant BX20220301). R.G.B. acknowledges financial support from the Severo Ochoa grant CEX2021-001131-S funded by MCIN/AEI/10.13039/501100011033. 
\end{acknowledgments}

\bibliographystyle{aasjournal.bst}

\bibliography{M81_cloudy_modeling}

\end{document}